\documentclass[
    aip,
    apl,
    amsmath,amssymb,
    reprint,
]{revtex4-1}

\usepackage{graphicx}

\usepackage[utf8]{inputenc}
\usepackage[T1]{fontenc}
\usepackage{dcolumn}
\usepackage{bm}
\usepackage{mathptmx}
\usepackage{multirow}
\usepackage{soul}
\usepackage[version=4]{mhchem}
\usepackage{float}
\usepackage{etoolbox}
\usepackage{anyfontsize}
\usepackage{titlesec}
\usepackage{hyperref}
\hypersetup{
    colorlinks=true,
    linkcolor=blue,
    citecolor=blue,
    urlcolor=blue
}
\renewcommand{\href}[2]{#2} 
\titleformat{\section}{\large\bfseries}{Section \thesection}{1em}{}%
\makeatletter
\def\@email#1#2{%
 \endgroup
 \patchcmd{\titleblock@produce}
  {\frontmatter@RRAPformat}
  {\frontmatter@RRAPformat{\produce@RRAP{*#1\href{mailto:#2}{#2}}}\frontmatter@RRAPformat}
  {}{}
}%
\makeatother
\newcommand{\beginsupplement}{%
    \setcounter{table}{0}
    \renewcommand{\thetable}{S\arabic{table}}%
    \setcounter{figure}{0}
    \renewcommand{\thefigure}{S\arabic{figure}}%
    \setcounter{section}{0}
    \renewcommand{\thesection}{S\arabic{section}}%
    \setcounter{equation}{0}
    \renewcommand{\theequation}{S\arabic{equation}}%
}

\begin{document}

\title{Gate tunable spin-charge interconversion in a graphene/ReS$_{2}$ heterostructure up to room temperature}

\author{Eoin Dolan}
\affiliation{CIC nanoGUNE BRTA, 20018 Donostia-San Sebastian, Basque Country, Spain}
\affiliation{Departamento de Polímeros y Materiales Avanzados: Física, Química y Tecnología, University of the Basque Country (UPV/EHU), 20018 Donostia-San Sebastian, Basque Country, Spain}

\author{Zhendong Chi}
\affiliation{CIC nanoGUNE BRTA, 20018 Donostia-San Sebastian, Basque Country, Spain}

\author{Haozhe Yang}
\affiliation{CIC nanoGUNE BRTA, 20018 Donostia-San Sebastian, Basque Country, Spain}
\affiliation{Fert Beijing Institute, MIIT Key Laboratory of Spintronics, School of Integrated Circuit Science and Engineering, Beihang University, China}

\author{Luis E. Hueso}
\affiliation{CIC nanoGUNE BRTA, 20018 Donostia-San Sebastian, Basque Country, Spain}
\affiliation{IKERBASQUE, Basque Foundation for Science, 48009 Bilbao, Basque Country, Spain}

\author{Fèlix Casanova$^{*}$}
\affiliation{CIC nanoGUNE BRTA, 20018 Donostia-San Sebastian, Basque Country, Spain}
\affiliation{IKERBASQUE, Basque Foundation for Science, 48009 Bilbao, Basque Country, Spain}

\begin{abstract}
($^{*}$Author to whom correspondence should be addressed: f.casanova@nanogune.eu)\\

Graphene is a material with great potential in the field of spintronics, combining good conductivity with low spin--orbit coupling (SOC), which allows for the transport of spin currents over long distances. However, this lack of SOC also limits the capacity for manipulating spin current. A key strategy to address this limitation is to induce SOC in graphene via proximity to other two-dimensional (2D) materials. Such proximity-induced SOC can enable spin--charge interconversion (SCI) in graphene, with potential applications in next-generation logic devices. Here, we place graphene in close proximity to the room-temperature ferroelectric candidate ReS$_\mathrm{2}$, inducing SCI for both in-plane and out-of-plane polarized spin current. We attribute the SCI for in-plane polarized current to either the Rashba--Edelstein effect (REE) or the unconventional spin Hall effect (SHE) at the graphene/ReS$_\mathrm{2}$ interface, and the SCI for out-of-plane polarized current to either the conventional SHE in the proximitised graphene, or the unconventional SHE in the bulk of the ReS$_\mathrm{2}$. SCI due to in-plane spin is characterised over a wide range of temperature, up to 300\,K and a range of gate voltages.
\end{abstract}

\maketitle

Graphene is of great interest for spintronic devices, due to its stability, conductivity, and an exceptionally long spin diffusion length due to its all-carbon composition.~\cite{avsar_colloquium_2020, han_graphene_2014, Zhou2024-tj} This spin diffusion length is on the order of hundreds of nanometers, even at room temperature and for relatively imperfect graphene. At low temperature, and in a pristine state, the diffusion length can be even longer, on the order of tens of microns.~\cite{ingla-aynes_24_2015, drogeler_spin_2016} However, this low spin-orbit coupling (SOC) in graphene is fundamentally incompatible with the manipulation of spin currents and the conversion to charge current (spin-charge interconversion, SCI). SCI is a key property for the realization of magnetoelectric spin-orbit (MESO) logic,~\cite{manipatruni_scalable_2019,pham_spinorbit_2020, vaz_voltage-based_2024} ferroelectric spin-orbit (FESO) logic,~\cite{noel_non-volatile_2020, Varotto2021-fo} or for magnetic random access memory (MRAM).~\cite{yang_two-dimensional_2022, lin_spin_2013} One possible way to achieve the desired SCI in graphene is by combining it with other materials with strong SOC, thus inducing SOC in the graphene by proximity.~\cite{weeks_engineering_2011, gmitra_spin-orbit_2013, calleja_spatial_2015, gmitra_trivial_2016, safeer_room-temperature_2019, safeer_large_2019, ghiasi_charge--spin_2019, khokhriakov_gate-tunable_2020, herling_gate_2020, safeer_spin_2020, benitez_tunable_2020, ingla-aynes_electrical_2021, ingla-aynes_omnidirectional_2022, safeer_reliability_2022, yang_gatetunable_2024, yang_twist-angle-tunable_2024, chi_control_2024}

This SCI in proximitized graphene occurs via either the spin Hall effect (SHE), which directly converts charge current to spin current (and spin to charge in the inverse effect),~\cite{sinova_spin_2015} and the Rashba-Edelstein effect (REE),~\cite{edelstein_spin_1990} which converts a charge current into a spin accumulation (and spin accumulation to charge current in the inverse effect). Both the SHE~\cite{safeer_room-temperature_2019, yang_gatetunable_2024} and REE~\cite{ghiasi_charge--spin_2019, khokhriakov_gate-tunable_2020} have been observed in different systems, or indeed occurring simultaneously in the same system.~\cite{benitez_tunable_2020, yang_twist-angle-tunable_2024} These effects allow for the generation and detection of spin currents without the use of ferromagnetic (FM) elements. In this letter, we demonstrate the induction of SCI in graphene via the proximity effect with ReS$_\mathrm{2}$, a transition metal dichalcogenide (TMD) with the 1T' phase reported to be ferroelectric.~\cite{wan_room-temperature_2022} ReS$_\mathrm{2}$ is an n-type semiconductor, meaning at positive gate voltages it can become conductive, allowing spin current to enter the ReS$_\mathrm{2}$. This permits further possibilities for SCI, with the potential to measure the SHE in the bulk of the 1T' phase of ReS$_\mathrm{2}$. \cite{ontoso_unconventional_2023, ingla-aynes_omnidirectional_2022, Camosi2022-yn, safeer_large_2019} Finally, in low-symmetry systems both the SHE and REE can have unconventional components, both in the bulk ReS$_\mathrm{2}$, and at the graphene/ReS$_\mathrm{2}$ interface\cite{ontoso_unconventional_2023, chi_control_2024}

\begin{figure}[t]
\includegraphics[width=\columnwidth]{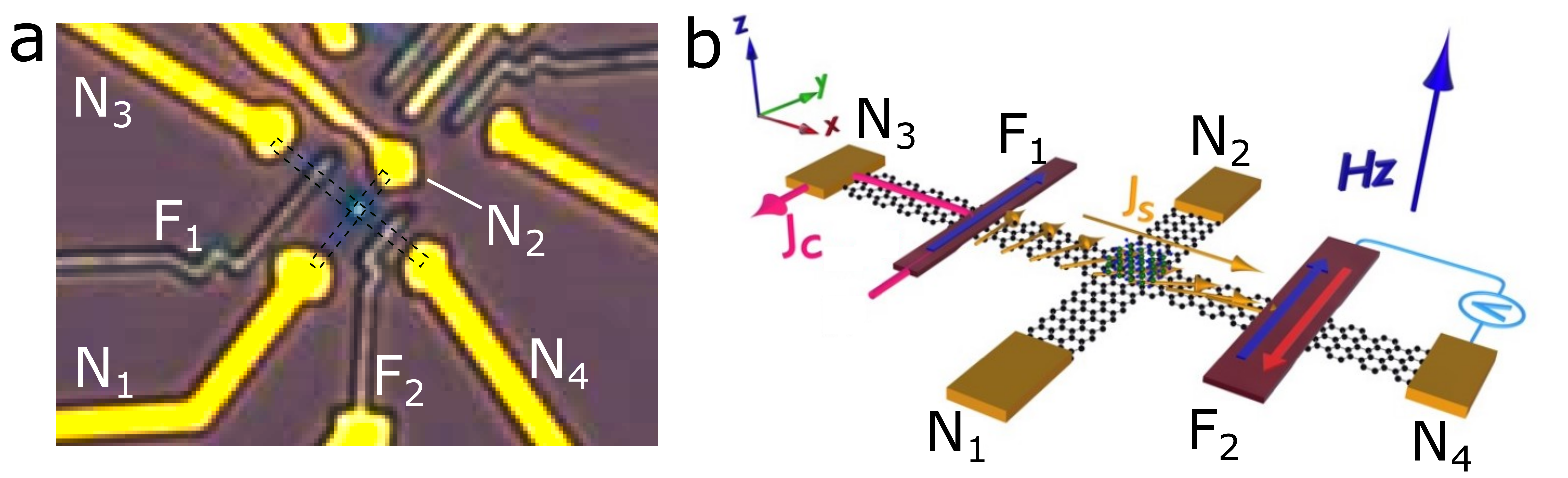}
\caption{\label{fig:epsart} Graphene/ReS$_\mathrm{2}$ device.
\textrm{(a)} An optical picture of the device showing non-magnetic electrodes for applying charge current ($\mathrm{N}_1$--$\mathrm{N}_4$), consisting of Pd/Au ($5\,\mathrm{nm}/35\,\mathrm{nm}$), and FM electrodes ($\mathrm{F}_1$, $\mathrm{F}_2$) for injecting/detecting spin current, consisting of $\mathrm{TiO}_x$/Co/Au ($0.3\,\mathrm{nm}/35\,\mathrm{nm}/15\,\mathrm{nm}$). The graphene is indicated with dashed lines in the shape of a cross, with the proximitizing ReS$_\mathrm{2}$ visible in the center.
\textrm{(b)} Device schematic in the spin transport configuration, where we inject a charge current ($\mathbf{J}_\mathrm{c}$) between $\mathrm{F}_1$ and $\mathrm{N}_3$. This causes spin current ($\mathbf{J}_\mathrm{s}$) to flow through the device until it reaches $\mathrm{F}_2$, where it is measured as a voltage change between $\mathrm{F}_2$ and $\mathrm{N}_4$. By application of an external magnetic field ($B_z$), the polarization of the spin current precesses in the $xy$-plane, modulating the non-local voltage as a function of $B_z$.}
\end{figure}

Our device consists of a few-layer graphene Hall bar (section S6), etched from a single flake of mechanically exfoliated graphene. The center of the Hall bar is proximitized with a flake of bulk 1T'-ReS$_\mathrm{2}$ (Fig.\,1a), placed via viscoelastic stamping. Lithographically patterned electrodes allow for the detection of charge current transverse to the channel, and FM cobalt (Co) electrodes enable the injection and detection of spin current into the channel. Full fabrication details are in Supplementary Material (section S10).

Measurements were performed by the application of a charge current ($\mathbf{J}_\mathrm{c}$) between an FM electrode and a non-magnetic electrode, with the current passing through the graphene ($\mathrm{F}_1$ to $\mathrm{N_3}$, Fig.\,1a). Once in the graphene channel, the spin current ($\mathbf{J}_\mathrm{s}$) diffuses away from $\mathrm{F}_1$ in both directions. It decays over a range of hundreds of nanometers, quantified by the spin diffusion length, $\lambda_\mathrm{s}$, which is affected by both gate voltage and temperature. $\mathbf{J}_\mathrm{s}$ can be detected via the second FM as a spin accumulation to probe spin transport, or via SCI. The voltage is measured in a non-local configuration, either from $\mathrm{F}_2$ to $\mathrm{N_4}$, to probe spin transport, or from $\mathrm{N}_1$ to $\mathrm{N_2}$, to probe SCI (Fig.\,1a). We report the non-local resistance ($R_\mathrm{NL}$), obtained by dividing the measured voltage ($V_\mathrm{NL}$) by the applied current ($\mathbf{J}_\mathrm{c}$). We measure $R_\mathrm{NL}$ as a function of the applied magnetic field, either in-plane ($B_x$) or out-of-plane ($B_z$), allowing us to study SCI due to out-of-plane ($\hat{\mathbf{s}}_z$) or in-plane ($\hat{\mathbf{s}}_x$) polarized spin current. The spin transport and SCI properties in the graphene are modulated by the application of back-gate voltage ($V_\mathrm{g}$), as well as temperature.

\begin{figure}[t]
    \centering
    \includegraphics[width=\columnwidth]{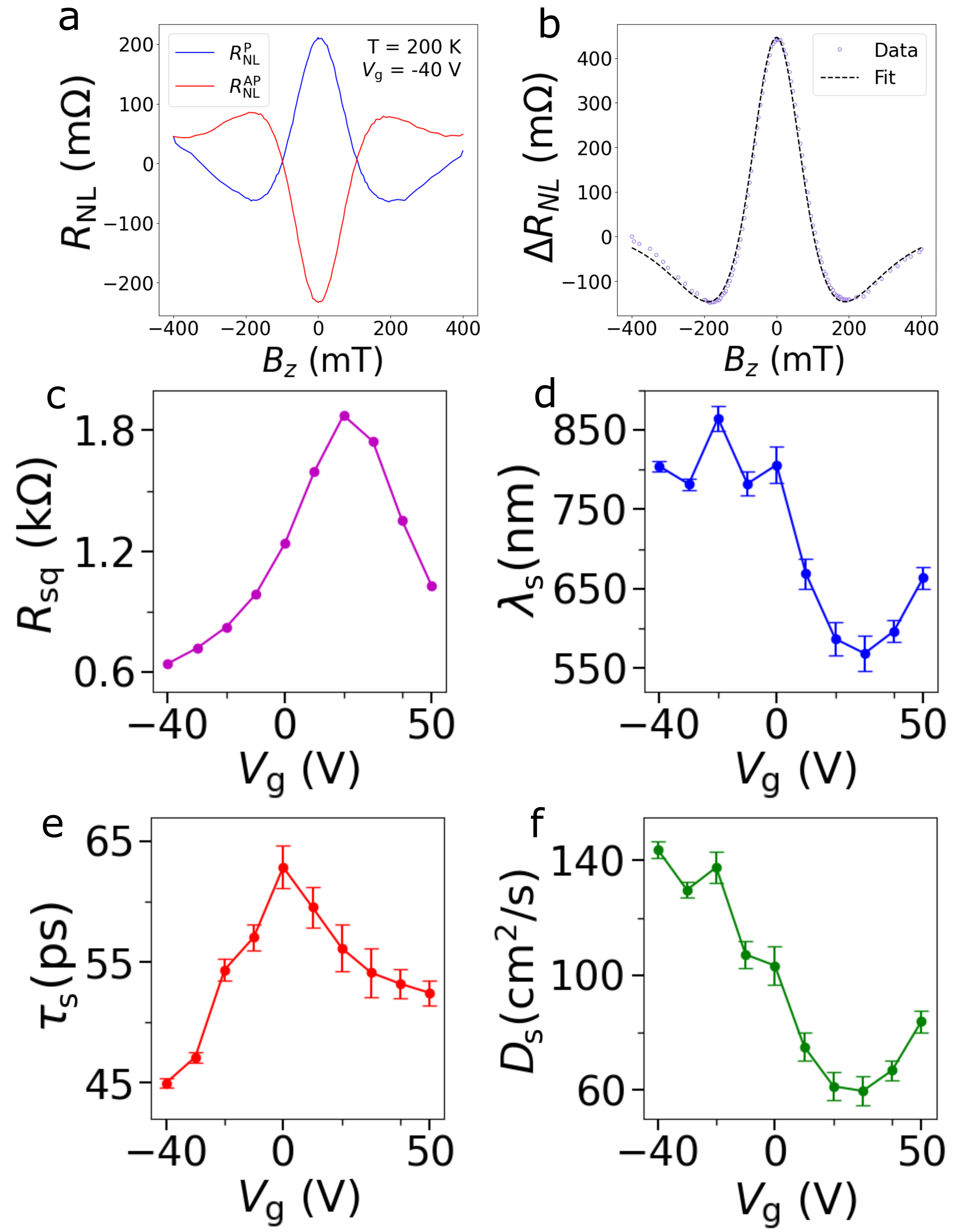}%
    \caption{\label{fig:epsart2}
    Spin transport in proximitized graphene.
    \textrm{(a)} Non-local resistance measured as a function of $B_z$ according to the configuration shown in Fig.\,1b. Two measurements are performed, corresponding to the initial relative magnetizations of $\mathrm{F}_1$ and $\mathrm{F}_2$, $R_\mathrm{NL}^\mathrm{P}$ and $R_\mathrm{NL}^\mathrm{AP}$. (b) The spin precession (Hanle) signal, obtained as the difference between $R_\mathrm{NL}^\mathrm{P}$ and $R_\mathrm{NL}^\mathrm{AP}$, fit to a one-dimensional spin-diffusion equation (Eq.\,\ref{eq:DeltaRNL}).
    \textrm{(c)} Square resistance of the region between $\mathrm{F}_1$ and $\mathrm{F}_2$, measured in a four-point configuration.
    \textrm{(d--f)} Spin-transport parameters of proximitized graphene extracted from the fitting in (b), showing the spin-diffusion length (d), spin lifetime (e), and spin-diffusion constant (f) at $200\,\mathrm{K}$ as a function of back-$V_\mathrm{g}$.
    }
\end{figure}

Before measuring SCI, we separately characterize the spin transport properties of the graphene/ReS$_\mathrm{2}$ device. This is done using the measurement configuration in Fig.\,1b, with $\mathbf{J}_\mathrm{c}=40\,\mu\mathrm{A}$. We apply $B_z$ to cause spin precession. Since the FM detector is sensitive to the $\hat{y}$ component of the spin, at zero field we see a maximum $R_\mathrm{NL}$, due to the accumulation of spin at the graphene/$\mathrm{F}_2$ interface, when $\mathrm{F}_1$ and $\mathrm{F}_2$ are magnetized in the same direction [Parallel ($R_\mathrm{NL}^\mathrm{P}$), Fig.\,2a], or a minimum when they are magnetized in opposite directions [Anti-Parallel ($R_\mathrm{NL}^\mathrm{AP}$), Fig.\,2a]. In principle, both Hanle curves in Fig.\,2a contain the same information; however, each one can also contain artifacts that do not come from spin precession, but rather from stray charge current, or pulling of the electrode magnetization by the external field, giving a component of spin oriented along $\pm\hat{z}$. To remove these, we subtract the two signals (Fig.\,2b), and fit with:~\cite{safeer_room-temperature_2019}

\begin{equation}
\Delta R_\mathrm{NL}
= \frac{P^2 \cos^2(\beta)\, R_\mathrm{sq} \,\lambda_\mathrm{s}}{w}
  \,\operatorname{Re} \Biggl\{
    \frac{e^{-\frac{L}{\lambda_\mathrm{s}}\,\Omega}}{\Omega}
  \Biggr\},
\label{eq:DeltaRNL}
\end{equation}

where $\Delta R_\mathrm{NL} = (R_\mathrm{NL}^\mathrm{P} - R_\mathrm{NL}^\mathrm{AP})$ is the spin precession (Hanle) signal. From this fitting, we extract the spin transport parameters of the system: the spin lifetime ($\tau_\mathrm{s}$), spin diffusion coefficient ($D_\mathrm{s}$), and spin diffusion length ($\lambda_\mathrm{s} = \sqrt{D_\mathrm{s} \tau_\mathrm{s}}$).

We also extract $P$, the interface spin polarization of the FM contact (see section S2). $\beta$ is the angle of the Co magnetization with respect to the easy axis (see sections S7 and S8 for details), $R_\mathrm{sq}$ is the square resistance of graphene (Fig.\,2c), $w$ is the channel width, $L$ is the channel length, and $\Omega = \sqrt{1 - i \omega \tau_\mathrm{s}}$, where $\omega = \frac{g \mu_\mathrm{B} B_z}{\hbar}$ is the Larmor frequency, with $g$ being the Landé $g$-factor, $\mu_\mathrm{B}$ the Bohr magneton, and $\hbar$ the reduced Planck's constant.

Fig.\,2c shows $R_\mathrm{sq}$ as a function of $V_\mathrm{g}$, with the charge neutrality point (CNP) at $\sim 20\,\mathrm{V}$. This is measured in a four-point configuration, applying $\mathbf{J}_\mathrm{c}$ from $\mathrm{N_3}$ to $\mathrm{N_4}$ and measuring the resulting voltage between $\mathrm{F}_1$ and $\mathrm{F}_2$. The shift away from $0\,\mathrm{V}$ is due to doping of the graphene during the fabrication process.

\begin{figure}[t]
\includegraphics[width=\columnwidth]{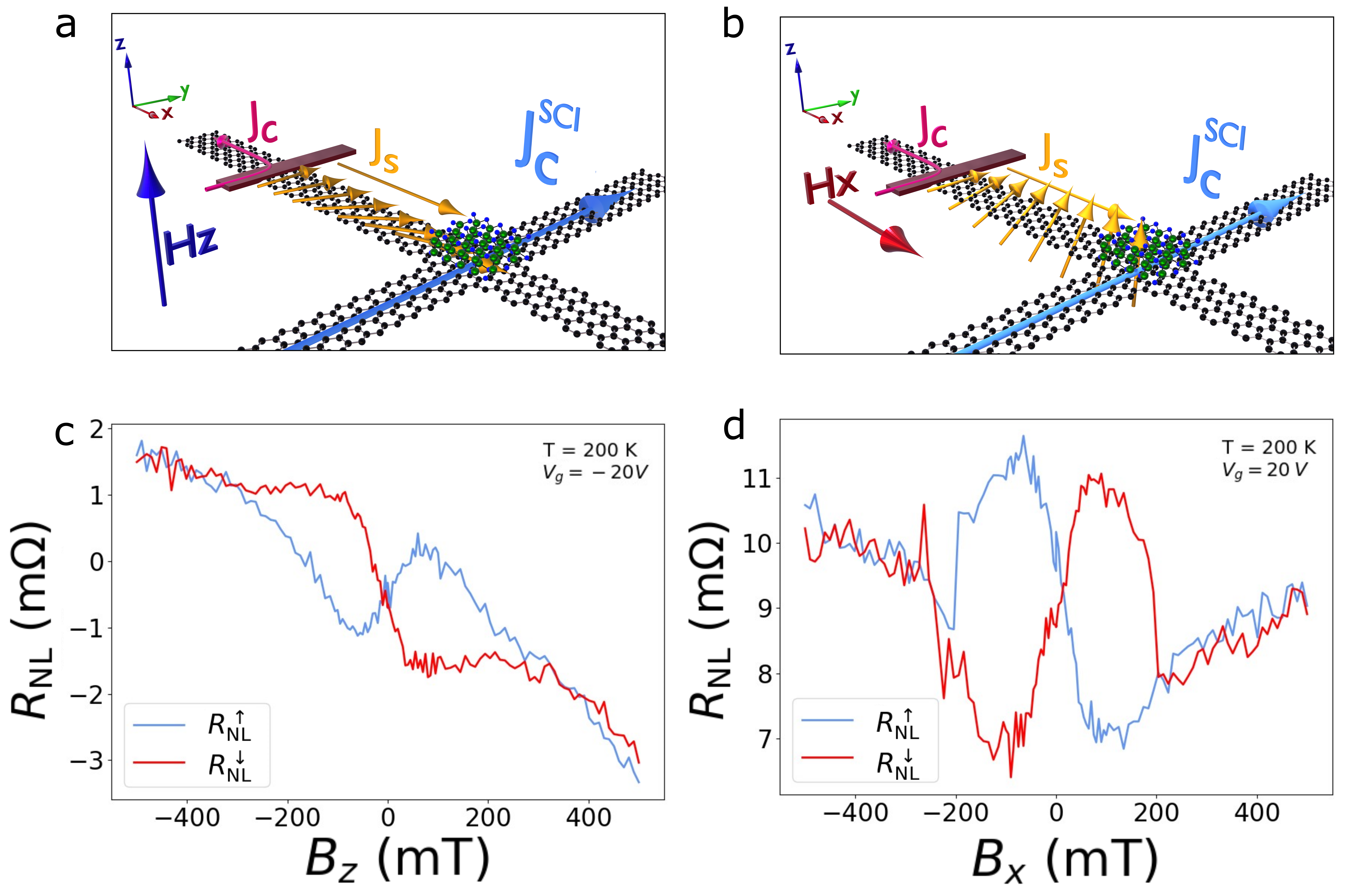}
\caption{\label{fig:epsart3} Spin-charge interconversion.
\textrm{(a)} Device design and measurement configuration for measuring SCI due to $\hat{\mathbf{s}}_x$ (magnetic field applied along $z$), showing applied charge current ($\mathbf{J}_\mathrm{c}$), the resultant spin current ($\mathbf{J}_\mathrm{s}$), and the charge current produced due to SCI in the proximitized region ($\mathbf{J}_\mathrm{c}^\mathrm{SCI}$).
\textrm{(b)} The same setup as (a), but now with the field applied along $x$, giving SCI due to $\hat{\mathbf{s}}_z$.
\textrm{(c)} Non-local resistance due to SCI from $\hat{\mathbf{s}}_x$, using the configuration shown in (a). Two measurements are performed, one with the injecting electrode magnetized along $+y$ ($R_\mathrm{NL}^\uparrow$), and the other with it magnetized along $-y$ ($R_\mathrm{NL}^\downarrow$).
\textrm{(d)} Non-local resistance due to SCI, measured according to the configuration in (b), with measurements performed for the injecting electrode initially magnetized along $+y$ and $-y$, as in (c).}
\end{figure}

Fixing $R_\mathrm{sq}$ and $\beta$, we fit the spin transport data using Eq.\,\ref{eq:DeltaRNL} with $\lambda_\mathrm{s}, D_\mathrm{s}, \tau_\mathrm{s}$ and $P$ as fitting parameters. These are plotted over a range of $V_\mathrm{g}$ (Fig.\,2d--f and Fig.\,S2).These measurements contain contributions from both pristine graphene and graphene proximitized with ReS$_\mathrm{2}$ (Fig.\,1b); the fitting parameters are therefore effective values for the whole system. Given the induction of SOC in the proximitized region, $\lambda_\mathrm{s}$ in this region is likely to be much shorter than our effective value, and the value in the pristine graphene somewhat longer. \cite{benitez_tunable_2020, Savero_Torres2017-qk}Additionally, the value of $\lambda_\mathrm{s}$ can be different for spin polarized in-plane compared to out-of-plane,\cite{Ghiasi2017-fa, Benitez2018-uc, Sierra2025-tc} although we cannot separate the two values and we use here an effective value. This can have a small effect on the extracted SCI efficiency (see section S9), which we fit based on the assumption of homogenous properties throughout the geometry, but without the possibility of doing the same measurement in the pristine graphene we cannot quantify the magnitude of the effect. 

Next, we characterize SCI in our device. SCI due to $\hat{\mathbf{s}}_z$ and $\hat{\mathbf{s}}_x$ are measured with the same electrical configuration, applying $\mathbf{J}_\mathrm{c}=40\,\mu\mathrm{A}$ between $\mathrm{F}_1$ and $\mathrm{N_3}$, while measuring voltage between $\mathrm{N_1}$ and $\mathrm{N_2}$. If we assume the ReS$_\mathrm{2}$ to be fully insulating, then a spin current in the proximitized region, $\mathbf{J}_\mathrm{s}$, always flows along $\pm\hat{x}$ due to the device geometry. It can be converted to a charge current, which is always measured between $\mathrm{N_1}$ and $\mathrm{N_2}$ (along $\hat{y}$, Fig.\,3a,b). $\mathbf{J}_\mathrm{c}$ and $\mathbf{J}_\mathrm{s}$ are fixed, but by applying an external magnetic field, we can control the spin polarization in the proximitized region via precession. In this scenario, there are two possible mechanisms for SCI: the SHE and the REE. For the SHE, there is a cross-product relationship between charge current, spin polarization, and spin current: $\mathbf{J}_\mathrm{c}^\mathrm{SHE} \propto \mathbf{J}_\mathrm{s} \times \hat{\mathbf{s}}$. Therefore, our experiment is sensitive to current produced by the SHE when there is spin polarization along $\hat{z}$. For the REE, in this two-dimensional system, the requirement to generate $\mathbf{J}_\mathrm{c}^\mathrm{REE}$ along $\hat{y}$ is that the spin be polarized along $\hat{x}$. However, both effects can also have unconventional components, when the symmetry of the system is lowered, for example when the graphene and the TMD are twisted by an angle.\cite{ontoso_unconventional_2023, chi_control_2024}

If we allow for the ReS$_\mathrm{2}$ being conductive, then SCI can also occur due to bulk effects in the ReS$_\mathrm{2}$. For conductive ReS$_\mathrm{2}$, $\mathbf{J}_\mathrm{s}$ is along $\hat{z}$, as it flows upwards from the graphene. For $\hat{\mathbf{s}}_x$ spins, we meet the condition for the conventional SHE in ReS$_\mathrm{2}$, and for $\hat{\mathbf{s}}_z$ spins, we can have an unconventional SHE, allowed by the 1T' crystal structure.~\cite{yang_twist-angle-tunable_2024, Ontoso2023-kr, ontoso_unconventional_2023, MacNeill2017-rd} Both of these effects will produce $\mathbf{J}_\mathrm{c}$ along $\hat{y}$, and therefore be measurable in this setup. Which effects we can measure therefore depend on both the external field, setting the spin polarization direction, and the gate voltage, which controls the ReS$_\mathrm{2}$ conductivity.

The initial phase of the spin polarization is given by the injector magnetization, $\pm\hat{y}$, and switches the sign of the signal, as seen in the red/blue curves ($R_\mathrm{NL}^{\uparrow/\downarrow}$) in Fig.\,3c,d. We can set this magnetization using an applied field along $\pm\hat{y}$ before beginning the measurement. We define the spin-charge interconversion resistance, $R_\mathrm{SCI}$, as $R_\mathrm{NL}^\uparrow - R_\mathrm{NL}^\downarrow$, which removes contributions due to charge current and injector magnetization pulling that do not change when the spin polarization reverses.~\cite{safeer_reliability_2022} This $R_\mathrm{SCI}$ is then fitted to:~\cite{safeer_room-temperature_2019}

\begin{equation}
R_\mathrm{SCI}
= \frac{P \,\theta_\mathrm{SCI} \cos(\beta)\, R_\mathrm{sq} \,\lambda_\mathrm{s}}{w_\mathrm{cr}}
\, \operatorname{Im} \Biggl\{
\frac{ e^{-\frac{L}{\lambda_\mathrm{s}}\,\Omega} - e^{-\frac{(L + w_\mathrm{cr})}{\lambda_\mathrm{s}}\,\Omega} }
{\Omega}
\Biggr\},
\label{eq:DeltaRSCI}
\end{equation}

where $w_\mathrm{cr}$ is the proximitized channel width. 

The efficiency of spin-charge interconversion, $\theta_\mathrm{SCI}$ is defined as $\mathbf{J}_\mathrm{c}^\mathrm{SCI} / \mathbf{J}_\mathrm{s}$, and can be due to either SHE or REE (conventional or unconventional), and SCI is assumed to occur in the proximitised graphene. This approach allows us to extract both the effective spin transport parameters (similar to Fig.\,2) and the SCI efficiency. The error associated with using effective values for the entire region for SCI efficiency calculation is discussed in section S9.

The figure of merit often used to compare SCI in different systems is $\theta_\mathrm{SCI} \times \lambda_\mathrm{s}$. $\lambda_\mathrm{s}$ is therefore crucial in quantifying the SCI, and it can be difficult to separate the contributions of $\theta_\mathrm{SCI}$ and $\lambda_\mathrm{s}$ in the measured signal, since they appear as a product in Eq.\,\ref{eq:DeltaRSCI}.

In the following analysis of the gate dependence of $\hat{\mathbf{s}}_x$-polarized SCI, the values for both $\theta_\mathrm{SCI}$ and $\lambda_\mathrm{s}$ are extracted by fitting $R_\mathrm{SCI}$ in Fig.\,4a, using $\theta_\mathrm{SCI}$, $\lambda_\mathrm{s}$, and $\tau_\mathrm{s}$ as free parameters. $P$ is fixed based on the spin transport measurements at $200\,\mathrm{K}$ (Fig.\,2) and assumed to remain constant with temperature. This is necessary because $P$ appears as a product with $\theta_\mathrm{SCI}$, making it otherwise impossible to separate them. $P$ is an interface property and in principle should not change with $V_\mathrm{g}$; however, in practice there is a small dependence (see section S2) which is accounted for in the fitting.

\begin{figure}[h!]
\includegraphics[width=0.48\textwidth]{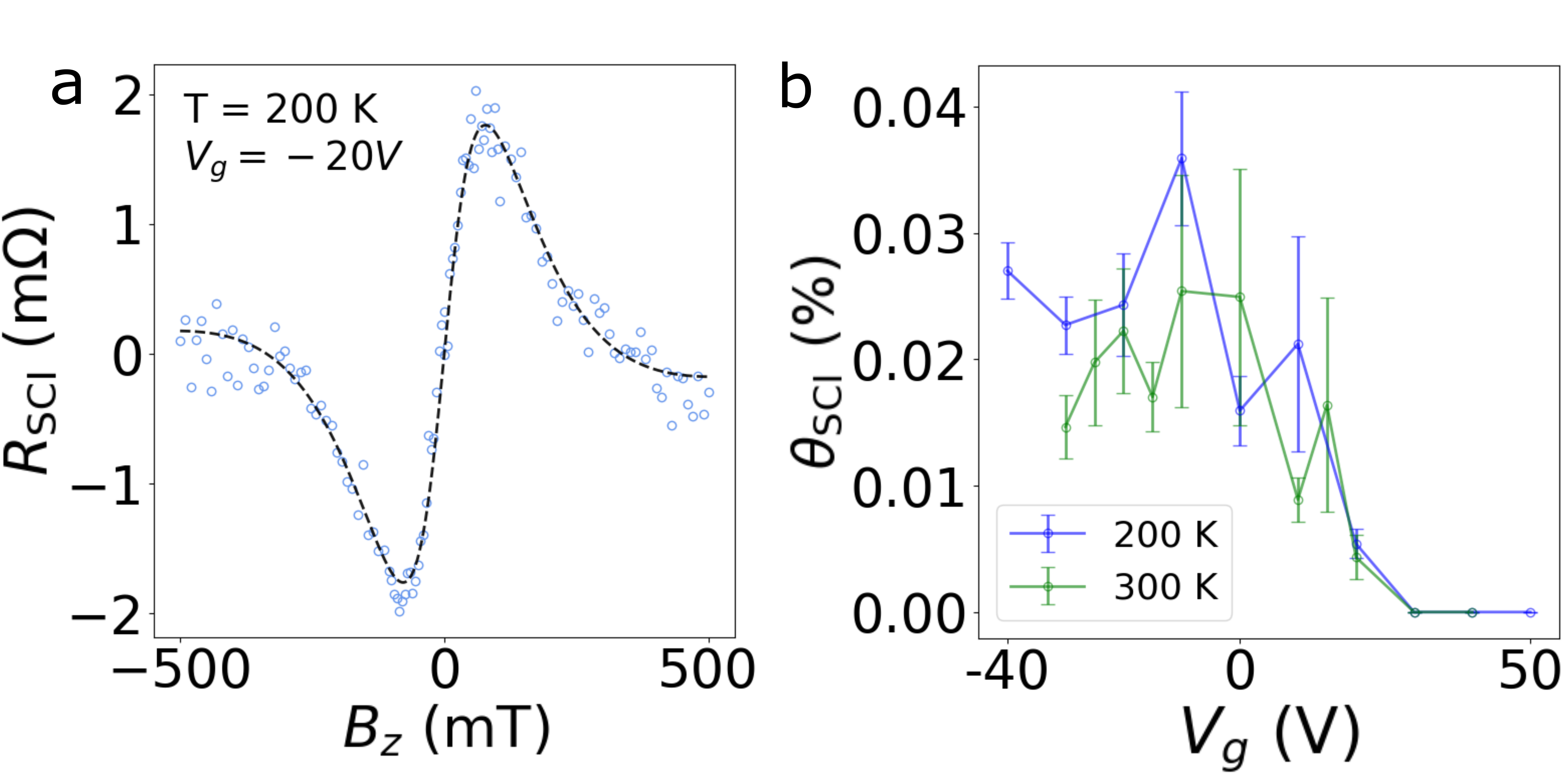}
\caption{\label{fig:epsart4} SCI efficiency.
\textrm{(a)} SCI resistance obtained by taking the difference in the two non-local resistance measurements in Fig.\,3c, isolating only the SCI contribution due to $\hat{\mathbf{s}}_x$ and removing any spurious artifacts. The dashed black line shows the fit to Eq.\,\ref{eq:DeltaRSCI}. \textrm{(b)} SCI efficiency, as a function of $V_\mathrm{g}$ at $200\,\mathrm{K}$ and $300\,\mathrm{K}$, obtained from the fitting in (a).}
\end{figure}

It is also possible to obtain the SCI efficiency parameters by extracting $\lambda_\mathrm{s}$ and $\tau_\mathrm{s}$, in addition to $P$, from the spin transport data (Fig.\,2), and then fitting the SCI data using only $\theta_\mathrm{SCI}$ as a free parameter. However, this approach is not generally applied in this work because we do not have spin transport values for all $V_\mathrm{g}$ and temperature points. For consistency, we therefore use only the SCI data to extract all parameters. Nevertheless, for comparison, we apply both methods at $200\,\mathrm{K}$ and $V_\mathrm{g} = -30\,\mathrm{V}$. Extracting all values from the SCI measurement yields $\lambda_\mathrm{s} = 880 \pm 84\,\mathrm{nm}$, $\theta_\mathrm{SCI} = 0.024 \pm 0.004\,\%$, giving $\lambda_\mathrm{s} \times \theta_\mathrm{SCI} = 0.21 \pm 0.04\,\mathrm{nm}$. Extracting $\lambda_\mathrm{s}$ and $\tau_\mathrm{s}$ from the spin transport data (Fig.\,2) and fitting the SCI data using only $\theta_\mathrm{SCI}$ as a free parameter gives $\lambda_\mathrm{s} = 781 \pm 7\,\mathrm{nm}$, $\theta_\mathrm{SCI} = 0.023 \pm 0.001\,\%$, resulting in $\lambda_\mathrm{s} \times \theta_\mathrm{SCI} = 0.18 \pm 0.01\,\mathrm{nm}$. Both approaches yield similar results, evidencing the consistency of the fitting. We perform a similar analysis for $\hat{\mathbf{s}}_z$-polarized SCI (Fig.\,3d), detailed in section S5. A summary of the fitted parameters is found in Table I. Note that although an $\hat{\mathbf{s}}_y$-polarized SCI would be allowed in a twisted heterostructure, \cite{yang_twist-angle-tunable_2024, lee_charge--spin_2022} it is not detected in our system ($R_\mathrm{SCI} = 0$ at $B_z=0$, Fig.\,4a).

\begin{table}[h!]
\centering
\caption{Comparison of SCI efficiencies.}
\label{tab:ree_she_values}
\resizebox{\columnwidth}{!}{%
\begin{tabular}{|c|c|c|c|c|c|}
\hline
Effect & T (K) & $V_\mathrm{g}$ (V) & $\theta_{\text{SCI}}$ (\%) & $\lambda_\mathrm{s}$ (nm) & $\lambda_\mathrm{s} \times \theta_{\text{SCI}}$ (nm) \\
\hline
SCI($\hat{\mathbf{s}}_x$) & 200 & -20 & $0.036 \pm 0.005$ & $744 \pm 52$ & $0.27 \pm 0.04$ \\ \cline{2-6}
& 300 & -20 & $0.022 \pm 0.005$ & $917 \pm 122$ & $0.20 \pm 0.05$ \\
\hline
SCI($\hat{\mathbf{s}}_z$) & 200 & 20  & $0.07 \pm 0.02$    & $586 \pm 21$  & $0.38 \pm 0.02$ \\
\hline
\end{tabular}%
}
\end{table}

The SCI for $\hat{\mathbf{s}}_x$ was measured over a range of $V_\mathrm{g}$ (Fig.\,4b) according to the scheme in Fig.\,3a, fitting with Eq.\,\ref{eq:DeltaRSCI} to extract $\theta_\mathrm{SCI}$. At negative $V_\mathrm{g}$ the efficiency does not vary greatly over a large range of $V_\mathrm{g}$ up to $0\,\mathrm{V}$. Between $0\,\mathrm{V}$ and $20\,\mathrm{V}$, the signal drops, and above $20\,\mathrm{V}$ we could not measure SCI. This region between $0\,\mathrm{V}$ and $20\,\mathrm{V}$ is where the conductivity of ReS$_\mathrm{2}$ increases rapidly (section S1). From the spin transport measurement in the same region, we see the value of $\lambda_\mathrm{s}$ (Fig.\,2d) fall by around 25\%. This could indicate absorption of spin current into ReS$_\mathrm{2}$. The flake of ReS$_\mathrm{2}$ does not completely cover the center of the Hall bar, leaving a path for some spin current to flow around the proximitized area. This explains why spin transport can still be measured even in the presence of absorption by ReS$_\mathrm{2}$. Such absorption has been observed previously and can be used to realize a spin transistor,~\cite{yan_two-dimensional_2016} or it may lead to the observation of SCI due to the conventional SHE in the bulk of the TMD material.~\cite{safeer_room-temperature_2019} Below $V_\mathrm{g} \approx 20\,\mathrm{V}$, as the ReS$_\mathrm{2}$ is highly insulating (section S1), we can attribute the SCI to the proximitized graphene, either due to the conventional REE or an unconventional SHE with $\mathbf{J}_\mathrm{s}$ parallel to $\hat{\mathbf{s}}_x$,\cite{chi_control_2024} since a negligible amount of $\mathbf{J}_\mathrm{s}$ can enter the ReS$_\mathrm{2}$. If the SCI due to $\hat{\mathbf{s}}_x$ were due to bulk effects, it would be expected to instead increase as the ReS$_\mathrm{2}$ becomes more conductive.

Above $V_\mathrm{g} = 20\,\mathrm{V}$, SCI from $\hat{\mathbf{s}}_x$ can no longer be resolved, even scanning up to $V_\mathrm{g} = 80\,\mathrm{V}$. $V_\mathrm{g} = 20\,\mathrm{V}$ also corresponds to the CNP in our system (Fig.\,2c). The fact that we cannot measure the SCI in the conductive ReS$_\mathrm{2}$ regime, is in agreement with the hypothesis of an effect in the proximitized graphene, due to either REE or unconventional SHE. Both effects would be expected to change sign at the CNP, and thus give zero SCI at this $V_\mathrm{g}$, although due to thermal broadening the real behaviour could be more complex. ~\cite{benitez_tunable_2020, garcia_spin_2018, lee_charge--spin_2022} The coincidence of the CNP and the onset of ReS$_\mathrm{2}$ conductivity makes it difficult to confirm which causes the SCI to fall to zero around $20\,\mathrm{V}$, but the failure of the SCI to re-emerge indicates that ReS$_\mathrm{2}$ can effectively absorb spin current and prevent SCI in the graphene at $V_\mathrm{g}$ above $20\,\mathrm{V}$. It should be noted that the coincidence of the CNP and the onset of the ReS$_\mathrm{2}$ conductivity is purely incidental, arising from doping of the graphene during fabrication.

SCI due to $\hat{\mathbf{s}}_z$ can be due to either the conventional SHE in the proximitized graphene or an unconventional SHE in the bulk ReS$_\mathrm{2}$. Our data for the $V_\mathrm{g}$ dependence of this SCI is very limited. We can clearly measure SCI at $+20\,\mathrm{V}$ (Fig.\,3b, see also section S5) from $100\,\mathrm{K}$ to $300\,\mathrm{K}$, and no SCI at $-20\,\mathrm{V}$ at all temperatures (S5), which is consistent with SCI due to the unconventional SHE in ReS$_\mathrm{2}$. However, the conductivity change of the ReS$_\mathrm{2}$ is not sharp, and it is possible that we can still measure the SHE in the proximitized graphene at this $V_\mathrm{g}$, as in practice due to thermal broadening the SHE is maximized at the CNP.~\cite{benitez_tunable_2020, garcia_spin_2017}

\begin{figure}[h!]
\includegraphics[width=0.48\textwidth]{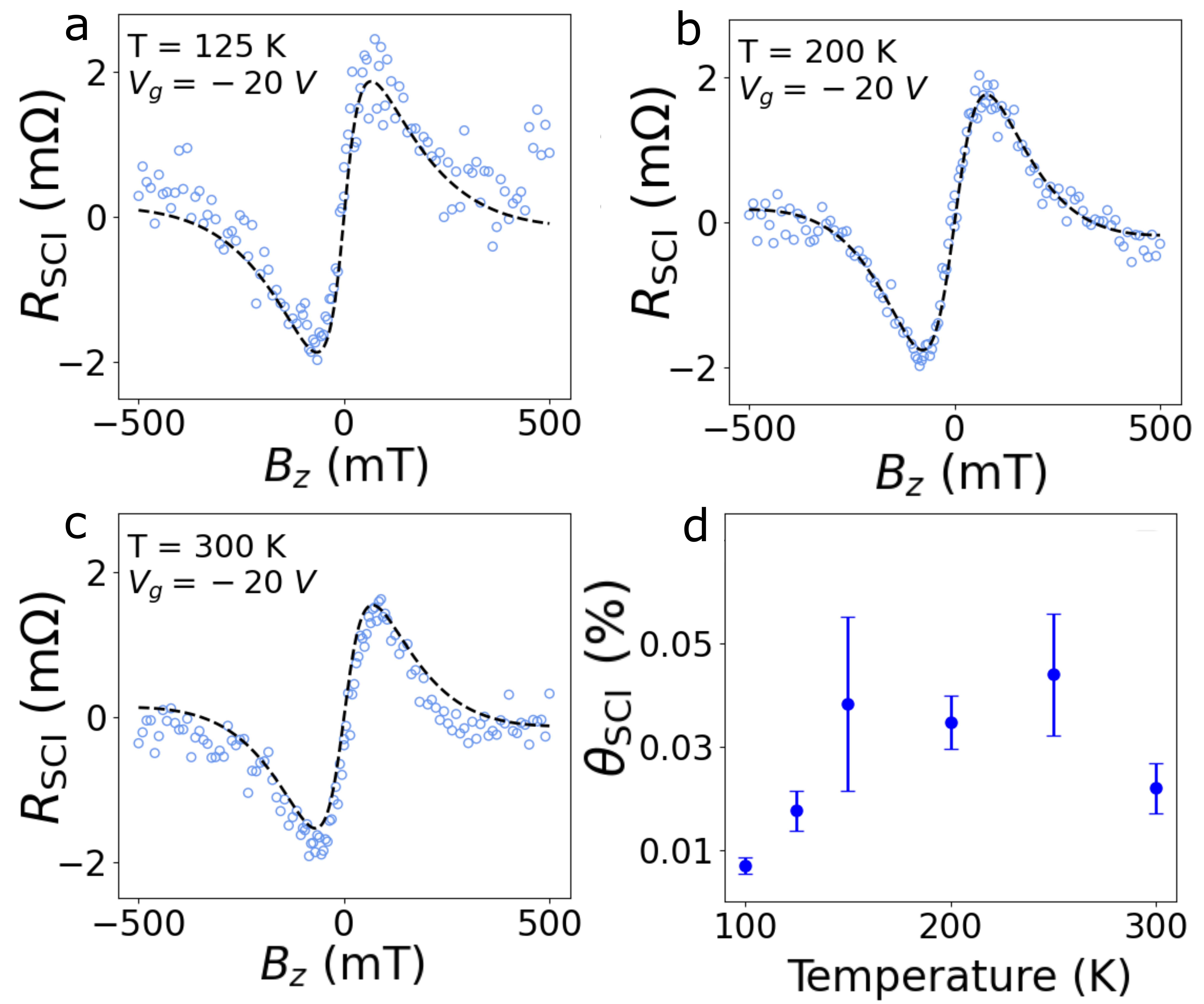}
\caption{\label{fig:epsart5} Non-local resistance due to SCI at different temperatures and $V_\mathrm{g}=-20\,\mathrm{V}$.
\textrm{(a)} $125\,\mathrm{K}$,
\textrm{(b)} $200\,\mathrm{K}$,
\textrm{(c)} $300\,\mathrm{K}$. The black dashed line shows the fit to Eq.\,\ref{eq:DeltaRSCI}. \textrm{(d)} SCI efficiency as a function of temperature.}
\end{figure}

Finally, $\hat{\mathbf{s}}_x$-polarized SCI was measured between $100\,\mathrm{K}$ and $300\,\mathrm{K}$ (Fig.\,5). All measurements were performed at $V_\mathrm{g}=-20\,\mathrm{V}$, except for the $100\,\mathrm{K}$ measurement. At $100\,\mathrm{K}$ and $V_\mathrm{g}=-20\,\mathrm{V}$, there was a very small signal, but an artifact in the measurement meant the data could not be fit well with Eq.\,\ref{eq:DeltaRSCI}. Therefore, the value for $\theta_\mathrm{SCI}$ was obtained from a measurement at $V_\mathrm{g}=-40\,\mathrm{V}$, where a good fit to Eq.\,\ref{eq:DeltaRSCI} was possible. The value of $\theta_\mathrm{SCI}$ at $100\,\mathrm{K}$ is assumed to be similar between $V_\mathrm{g} = -20\,\mathrm{V}$ and $V_\mathrm{g} = -40\,\mathrm{V}$ based on the data at $200\,\mathrm{K}$ and $300\,\mathrm{K}$ (Fig.\,4b). From this data (Fig. 5d), it appears that $\theta_\mathrm{SCI}$ reaches a maximum around 200\,K, but the reasons for this are not clear. When calculating simultaneously $\lambda_\mathrm{s}$, we see a similar but opposite trend (see section S4), such that the product of the two is nearly constant within the uncertainty. This apparent temperature maximum might therefore be a result of the correlation between $\theta_\mathrm{SCI}$ and $\lambda_\mathrm{s}$.

In conclusion, we demonstrate both in-plane and out-of-plane SCI up to room temperature in a graphene/ReS$_\mathrm{2}$ system. The in-plane SCI arises either due to the conventional REE induced in the graphene by the ReS$_\mathrm{2}$ or an unconventional SHE. We measure the gate dependence of this SCI, and note that it disappears above $20\,\mathrm{V}$. This is consistent with the SCI falling toward zero at the graphene CNP, where it is expected to switch sign. We propose that the SCI in the proximitized graphene does not re-emerge above the CNP with the opposite sign because around this same $V_\mathrm{g}$ the ReS$_\mathrm{2}$ becomes significantly more conductive, thus absorbing the spin current. In this region, we are able to measure the out-of-plane SCI which may arise due to either SHE in proximitized graphene or an unconventional SHE in the bulk of the 1T'-ReS$_\mathrm{2}$. This work establishes SCI in an unexplored graphene/TMD system, which can be tuned via $V_\mathrm{g}$. The ability to switch the spin symmetry of the SCI offers promising possibilities for the development of spintronic logic devices.

\begin{acknowledgments}
The authors acknowledge funding from MICIU/AEI/10.13039/501100011033 (Grant No. CEX2020-001038-M), from MICIU/AEI and ERDF/EU (Project No. PID2021-122511OB-I00), from the European Union's Horizon 2020 research and innovation programme under the Marie Skłodowska-Curie grant agreement No. 955671 and from the Valleytronics Intel Science and Technology Center. Z.C acknowledges funding from MICIU/AEI and the European Union NextGenerationEU/PRTR (Grant No. FJC2021-047257-I).
\end{acknowledgments}

\clearpage
\onecolumngrid

\begin{center}
\textbf{\huge Supplementary Material}
\end{center}

\beginsupplement
\fontsize{12}{14}\selectfont

\section{$\mathbf{ReS_2}$ conductivity}

\begin{figure}[H]
\centering
\includegraphics[width=0.6\linewidth]{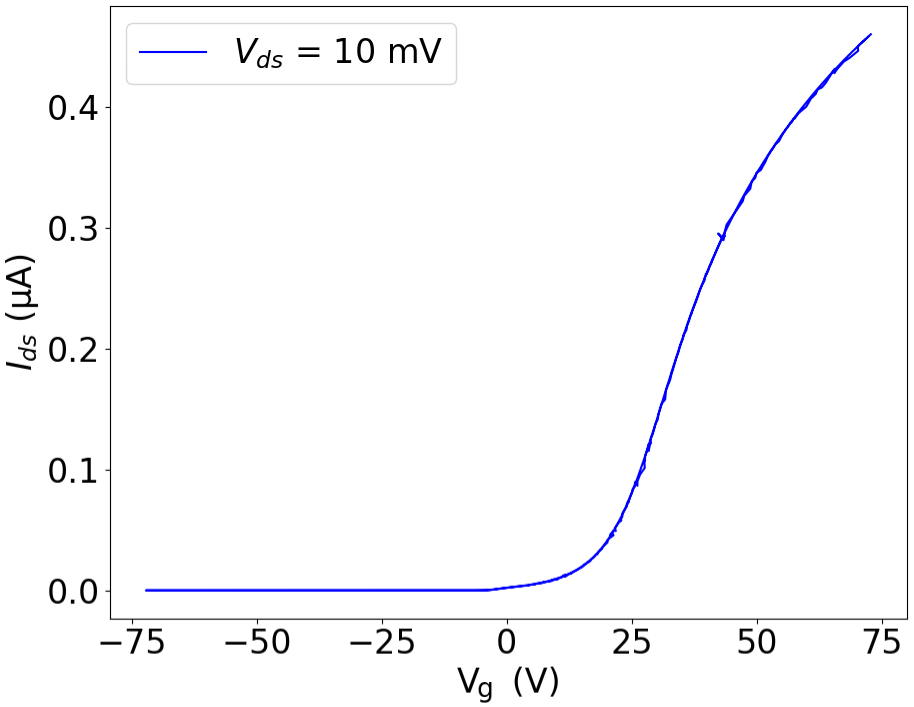}
\caption{Drain-source current ($I_\mathrm{ds}$) of a graphene/ReS\(_{2}\) system measured at a drain-source voltage ($V_\mathrm{ds}=10\,\mathrm{mV}$) as a function of gate voltage ($V_\mathrm{g}$). The data indicate how the system’s conductivity changes with gate bias.}
\label{fig:SI2}
\end{figure}

The conductivity of the graphene/ReS\(_{2}\) system is of key interest. In the device studied here (see Fig.~1a of the main text), the ReS\(_{2}\) flake is very small, making it impossible to measure its conductivity directly. Instead, we characterized a reference device fabricated from the same bulk flakes of graphene and ReS\(_{2}\). In this reference device, a cross was created by stamping a ReS\(_{2}\) flake onto a graphene flake. A constant drain-source voltage (\(\mathit{V}_{\mathrm{ds}}\)) of 10\,mV was applied across these two materials while sweeping the gate voltage (\(\mathit{V}_{\mathrm{g}}\)), and the resulting current (\(\mathit{I}_{\mathrm{ds}}\)) was recorded. Since ReS\(_{2}\) is an \(n\)-type semiconductor, it is expected to become more conductive at positive \(\mathit{V}_{\mathrm{g}}\). Consistent with this expectation, a sharp increase in the measured \(\mathit{I}_{\mathrm{ds}}\) (and thus the conductivity) is observed around \(\mathit{V}_{\mathrm{g}} = 20\)\,V.

\newpage
\section{Interface spin polarization of the ferromagnetic electrode as a function of \texorpdfstring{$V_\mathrm{g}$}{Vg}}
\begin{figure}[H]
\centering
\includegraphics[width=0.5\linewidth]{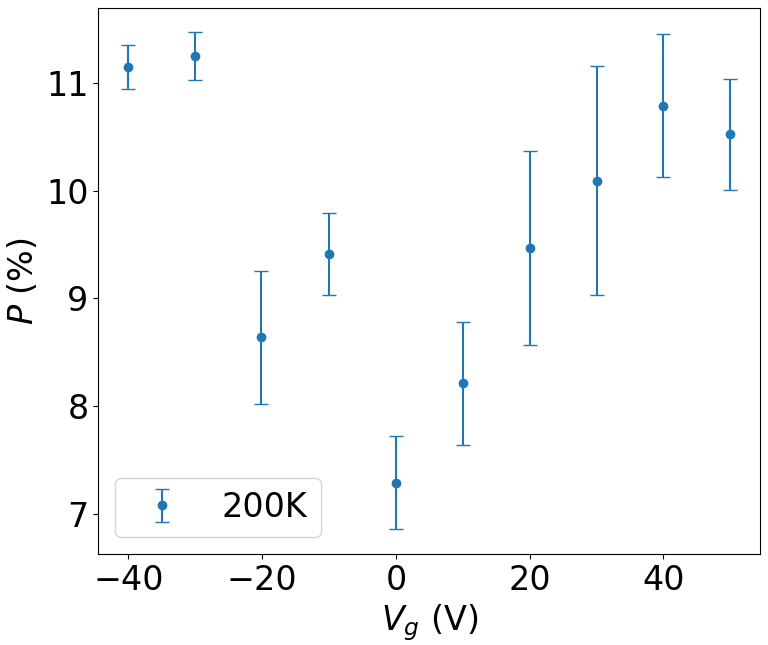}
\caption{Interface spin polarization (\(P\)) as a function of $V_\mathrm{g}$, extracted from the spin transport measurement at 200\,K shown in Fig. 2 of the main text.}
\label{fig:SI1}
\end{figure}

The interface spin polarization \(P\), appearing in both Eq.~1 and Eq.~2 of the main text, is fitted from the spin transport data at 200 K and summarized in Fig.~S2. The results for \(P\) as a function of $V_\mathrm{g}$ are shown here. Since spin transport measurements are not available for all temperatures and $V_\mathrm{g}$, \(P\) is assumed to be constant in temperature. When \(P\) is required at values of $V_\mathrm{g}$ not measured here, we use a linear interpolation of these data.

These data come from a measurement involving two interface polarizations - the injector and the detector. These are of equal importance, and the quantity \(P\) appearing in Eq.\ 1 of the main text is actually \(\sqrt{P_{\mathrm{inj}} \, P_{\mathrm{det}}}\). In contrast, the SCI resistance contains only one source of \(P\), namely the injector (Eq.\ 2). Therefore, to assign a value to \(P\) in Eq.\ 2 we need to assume that \(P_{\mathrm{inj}}\) and \(P_{\mathrm{det}}\) are the same. Although the TiO\textsubscript{x} barriers between the Co and graphene are grown at the same time and under identical conditions, their polarizations can still differ, and this is a key source of the uncertainty when fitting the SCI data for \(\theta_{\mathrm{SCI}}\).

\newpage
\section{\texorpdfstring{\(\theta_{\mathrm{SCI}} \times \lambda_{\mathrm{s}}\)}{theta SCI lambda s} product}

\begin{figure}[H]
\centering
\includegraphics[width=1\linewidth]{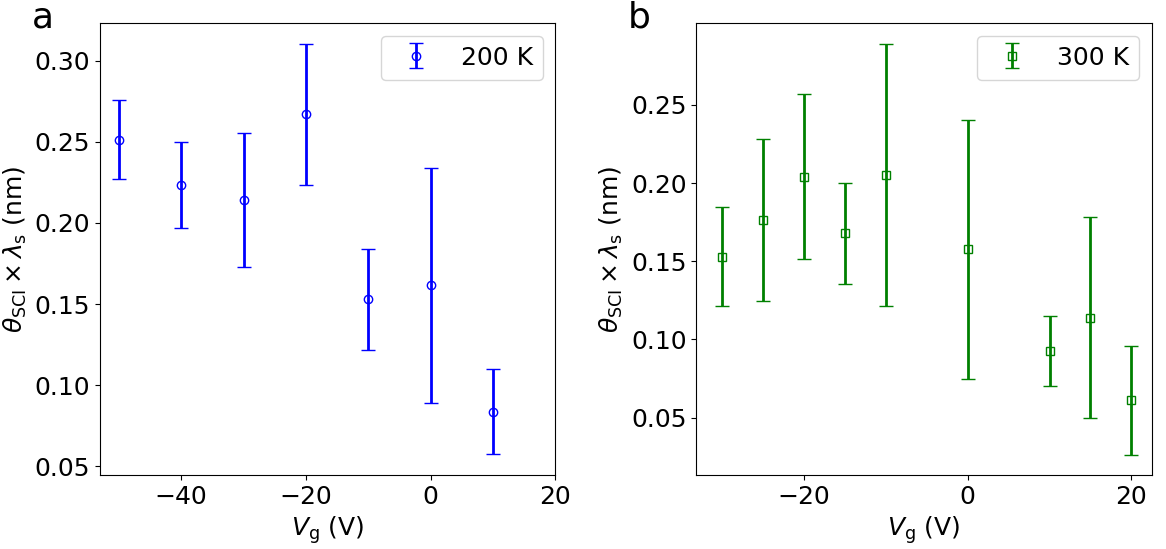}
\caption{Product of \(\theta_{\mathrm{SCI}}\) and \(\lambda_{\mathrm{s}}\), corresponding to the data in Fig.~4 of the main text}
\label{fig:SI3}
\end{figure}

In the main text, Fig.~4, we show \(\theta_{\mathrm{SCI}}\) for the system as a function of $V_\mathrm{g}$ at two temperatures. This comes from fitting the SCI data using \(\theta_{\mathrm{SCI}}\), \(\tau_\mathrm{s}\), and \(\lambda_{\mathrm{s}}\) as free parameters. From this same fitting, we can therefore also extract the \(\theta_{\mathrm{SCI}} \times \lambda_{\mathrm{s}}\) product, which is shown here.

\newpage

\section{Temperature dependence of $\theta_\mathrm{SCI}$ , $\lambda_\mathrm{s}$, and $\theta_\mathrm{SCI} \times \lambda_\mathrm{s}$}

As shown in the main text, there appears to be a maximum value of SCI efficiency around 200\,K. We show here a simultaneously extracted value for $\lambda_\mathrm{s}$, which shows an opposing trend, i.e. reaching a minimum around 200\,K. This is comparable to the trend seen in [\onlinecite{Zhou2024-tjAAA}], but reversed, showing a minimum rather than a maximum value for $\lambda_\mathrm{s}$ at 200\,K. However, when we look at the product $\theta_\mathrm{SCI} \times \lambda_\mathrm{s}$, we see that it is approximately constant across a range of temperature, meaning the ostensible trend in $\theta_\mathrm{SCI}$ could be due to a strong correlation between the two fitted values.

\begin{figure}[H]
\centering
\includegraphics[width=1\linewidth]{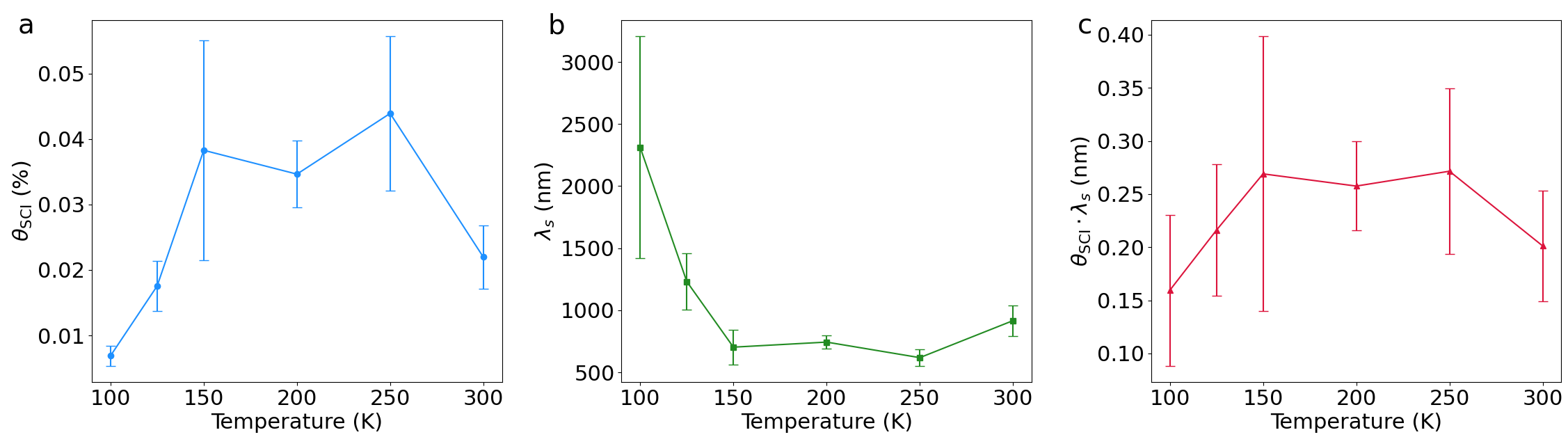}
\caption{Comparison of a) SCI efficiency, b) spin diffusion length, and c) the product $\theta_\mathrm{SCI} \times \lambda_\mathrm{s}$ extracted from SCI data at $V_\mathrm{g} = -20\,\mathrm{V}$ ($V_\mathrm{g} = -40\,\mathrm{V}$ for $100\,\mathrm{K}$).}

\label{fig:thetaandlambda}
\end{figure}

\newpage

\section{Out-of-plane SCI}
\label{sec:she_gate_dependence}

\begin{figure}[!ht]
\centering
\includegraphics[width=\linewidth]{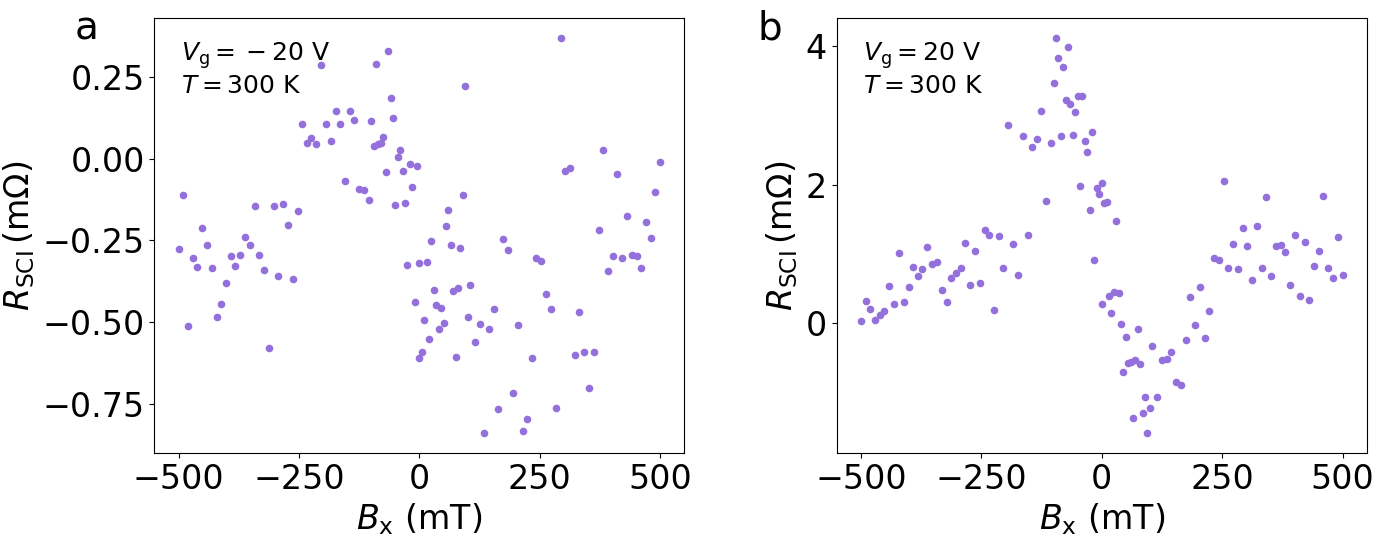}
\caption{SCI measurements at $300\,\mathrm{K}$. (a) At $V_\mathrm{g} = -20\,\mathrm{V}$, no SCI signal is observed. (b) At $V_\mathrm{g} = +20\,\mathrm{V}$, a clear signal is present.}
\label{fig:SHE_300K}
\end{figure}

\begin{figure}[!ht]
\centering
\includegraphics[width=\linewidth]{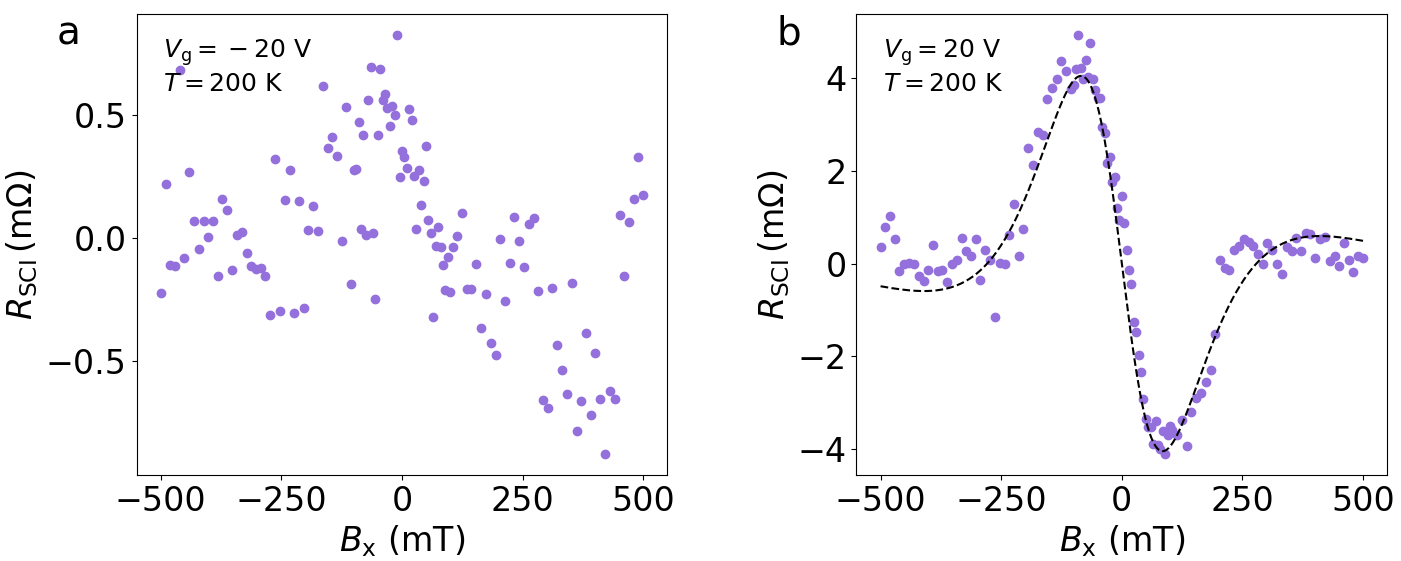}
\caption{SCI measurements at $200\,\mathrm{K}$. (a) Similar to the room temperature case, no signal is present at $V_\mathrm{g} = -20\,\mathrm{V}$. (b) SCI is measured at $V_\mathrm{g} = +20\,\mathrm{V}$ and fitted to the 1D analytical model.}
\label{fig:SHE_200K}
\end{figure}

\begin{figure}[!ht]
\centering
\includegraphics[width=\linewidth]{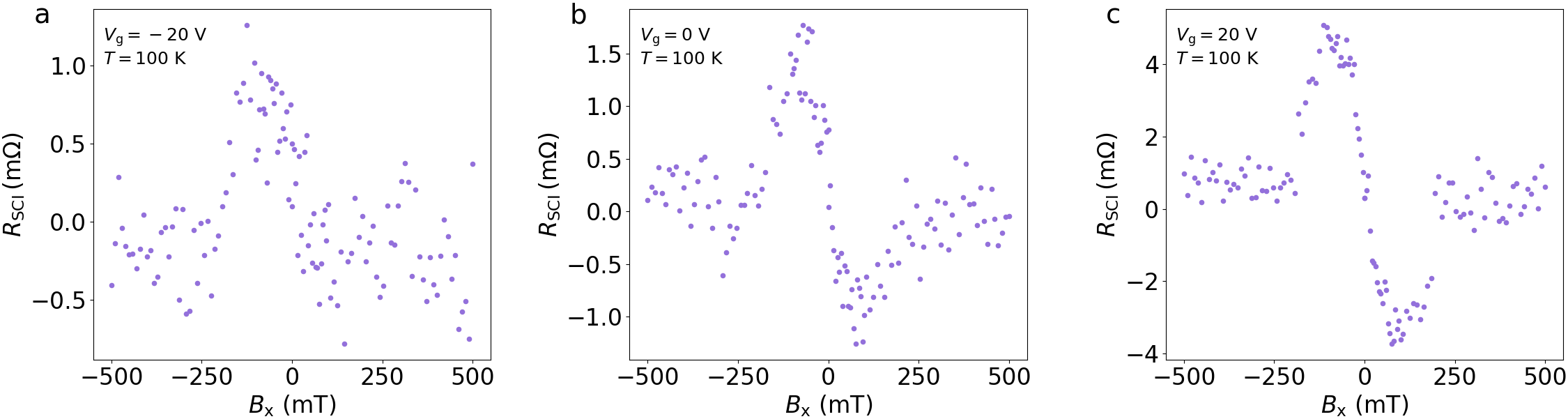}
\caption{SCI measurements at $100\,\mathrm{K}$. Panels show data at (a) $V_\mathrm{g} = -20\,\mathrm{V}$, (b) $V_\mathrm{g} = 0\,\mathrm{V}$, and (c) $V_\mathrm{g} = +20\,\mathrm{V}$.}
\label{fig:SHE_100K}
\end{figure}

To investigate the SCI due to out-of-plane spins, we measured the non-local signal as a function of $V_\mathrm{g}$. The data, presented in Figs.~\ref{fig:SHE_300K} to \ref{fig:SHE_100K}, consistently show that an SCI signal is resolved only at a positive gate voltage of $V_\mathrm{g} = +20\,\mathrm{V}$. At negative gate voltages ($V_\mathrm{g} = -20\,\mathrm{V}$), no clear signal is observed. At 100\,K, we can also measure a small signal at $V_\mathrm{g}=0\,V$ (Fig. \ref{fig:SHE_100K}b).

This gate voltage dependence is consistent with having SCI due to out-of-plane spins only when the ReS$_\mathrm{2}$ layer is conductive. This supports the hypothesis that the measured SCI arises from an unconventional SHE within the bulk of the ReS$_\mathrm{2}$ layer, rather than from a proximity-induced SHE in the graphene, which would likely be present at both positive and negative gate voltages. However, the conventional SHE in practice is expected to be largest near the CNP [\onlinecite{benitez_tunable_2020AAA}], so we do not exclude the possibility that the absence of a signal at $V_\mathrm{g} = -20\,\mathrm{V}$ is simply due to us being far from the CNP, and the SCI is due to SHE in the proximitized graphene.

This out-of-plane SCI data could not be uniquely fitted using Eq.~2 as the covariance between \(\theta_{\mathrm{SCI}}\) and \(\lambda_\mathrm{s}\) is too strong. At 200\,K, because we had independently measured \(\lambda_\mathrm{s}\) and \(\tau_\mathrm{s}\) from spin transport measurements, we could fix these values and fit the SCI data using only \(\theta_{\mathrm{SCI}}\), giving a good fit to the data (Fig. \ref{fig:SHE_200K}b). At other temperatures, we did not have the spin transport measurements to do this fitting. Therefore, we show the $R_\mathrm{SCI}$ data at 100\,K and 300\,K without any fitting.

\newpage

\section{Raman characterization of graphene}

\begin{figure}[H]
\centering
\includegraphics[width=0.6\linewidth]{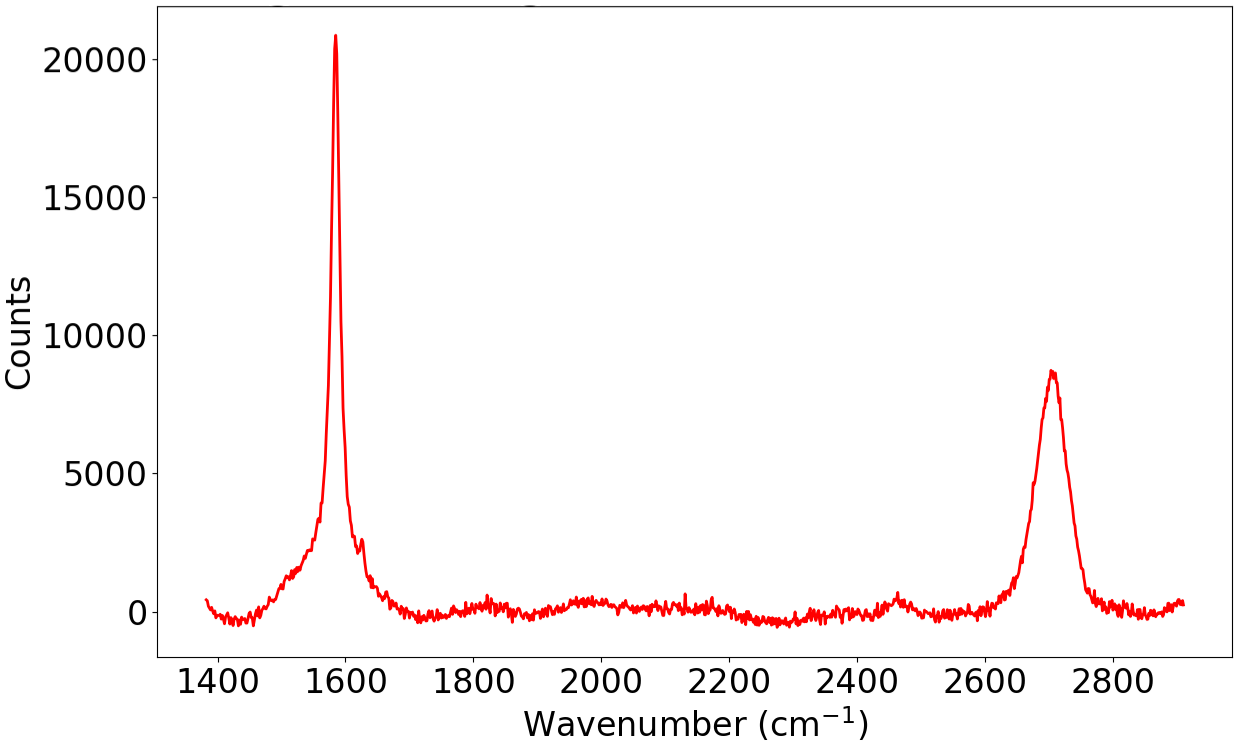}
\caption{Raman spectrum of the pristine part of the graphene flake, taken using a 532 nm laser.}
\label{fig:SI5}
\end{figure}

The graphene used in this work is few-layer (approximately 3 layers). This is judged from optical contrast during the exfoliation process. In principle, the thickness can be found using Raman spectroscopy, but this is not definitive outside of the monolayer case. The Raman spectrum of this graphene is consistent with 3-layer graphene, but this is not a definitive way to measure the thickness. In all analyses, we have considered the square resistance and can disregard the precise thickness.
\newpage

\section{Injector magnetization pulling with $B_x$}

\begin{figure}[H]
\centering
\includegraphics[width=1\linewidth]{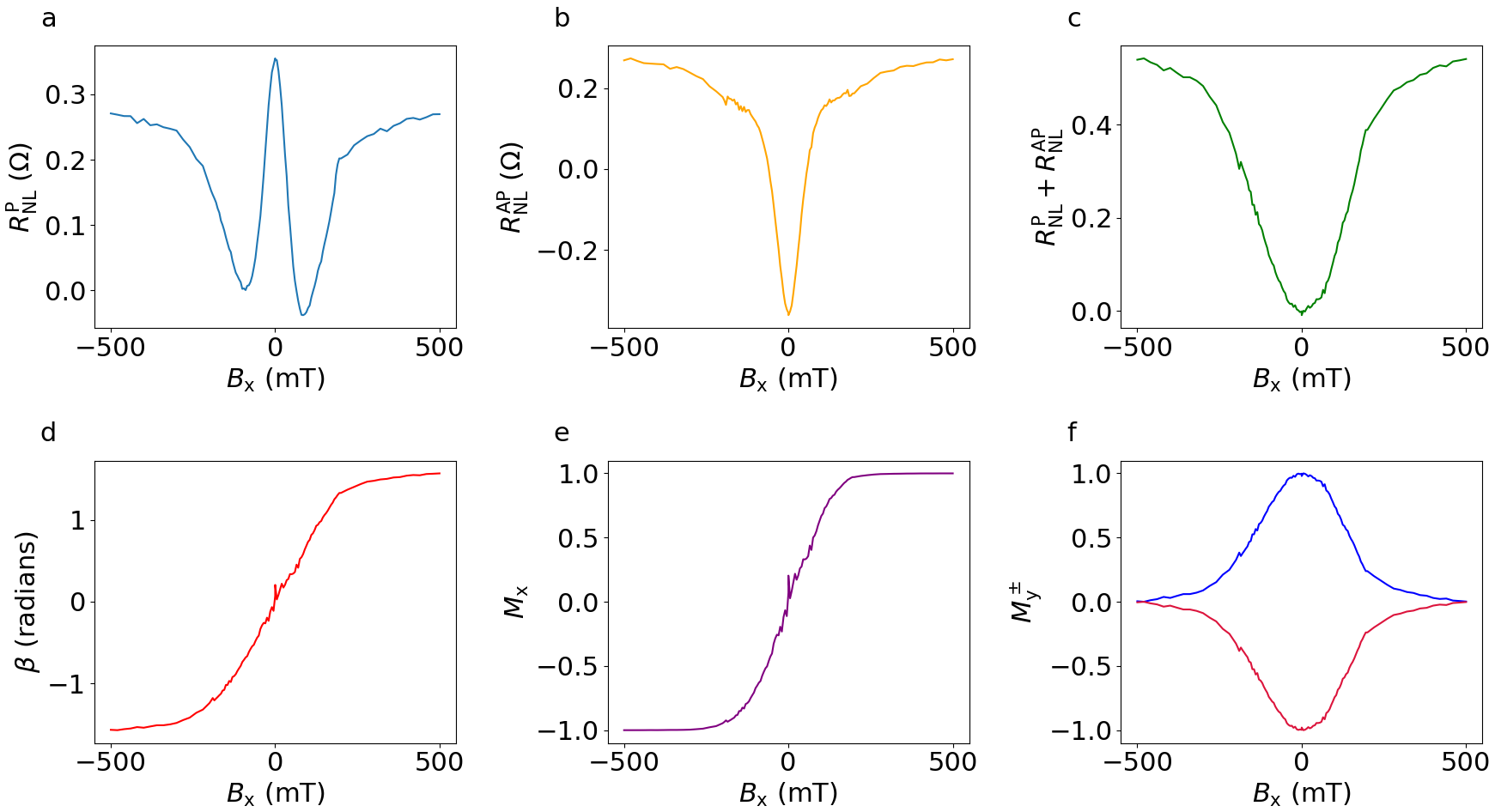}
\caption{Calculation of $\beta$ as used in Eq.~2 for the case of an in-plane field ($B_x$). The spin diffusion measurement described in Fig.~1b of the main text is shown, although in this case with $B_x$, with (a) parallel injector/detector magnetization, (b) anti-parallel configuration, (c) the sum of (a) and (b), and (d) the extracted value of $\beta$, the angle between the electrode magnetization and its easy axis. e,f) show the corresponding values of the unit magnetization vector, M, along x and y respectively}
\label{fig:SI6}
\end{figure}

For an applied in-plane magnetic field, the angle \(\beta\) is determined from a spin-diffusion measurement performed on a reference device, with a different graphene flake, but the same Co shape and thickness. In this measurement, the non-local resistance is recorded in the parallel \((P)\) configuration~(Fig. \ref{fig:SI6}a) and the anti-parallel \((AP)\) configuration~(Fig. \ref{fig:SI6}b) as before, however instead of taking the difference to isolate the precession component as in Fig. 2 of the main text, we add the two signals together~(Fig. \ref{fig:SI6}c) effectively cancelling any spin-precession contribution, because the precession term (see the exponential in Eq.1 of the main text) changes sign between \(P\) and \(AP\). As a result, we obtain \(R_{\mathrm{NL}}^{\mathrm{sum}} = (R_{\mathrm{NL}}^{\mathrm{P}}+R_{\mathrm{NL}}^{\mathrm{AP}}) \propto \cos^2(\beta)\), allowing us to extract \(\beta\), noting that the \(\pm\) saturation values of \(\beta\) are \(\pm \tfrac{\pi}{2}\) (Fig. \ref{fig:SI6}d). Figs. \ref{fig:SI6}e,f show directly the magnitude of the unit magnetization vectors along x (which is independent of initial magnetization), and along y which can be $\pm$1 at zero applied field depending on the initial magnetization.

\newpage
\section{Injector magnetization pulling with $B_z$}

\begin{figure}[H]
\centering
\includegraphics[width=0.8\linewidth]{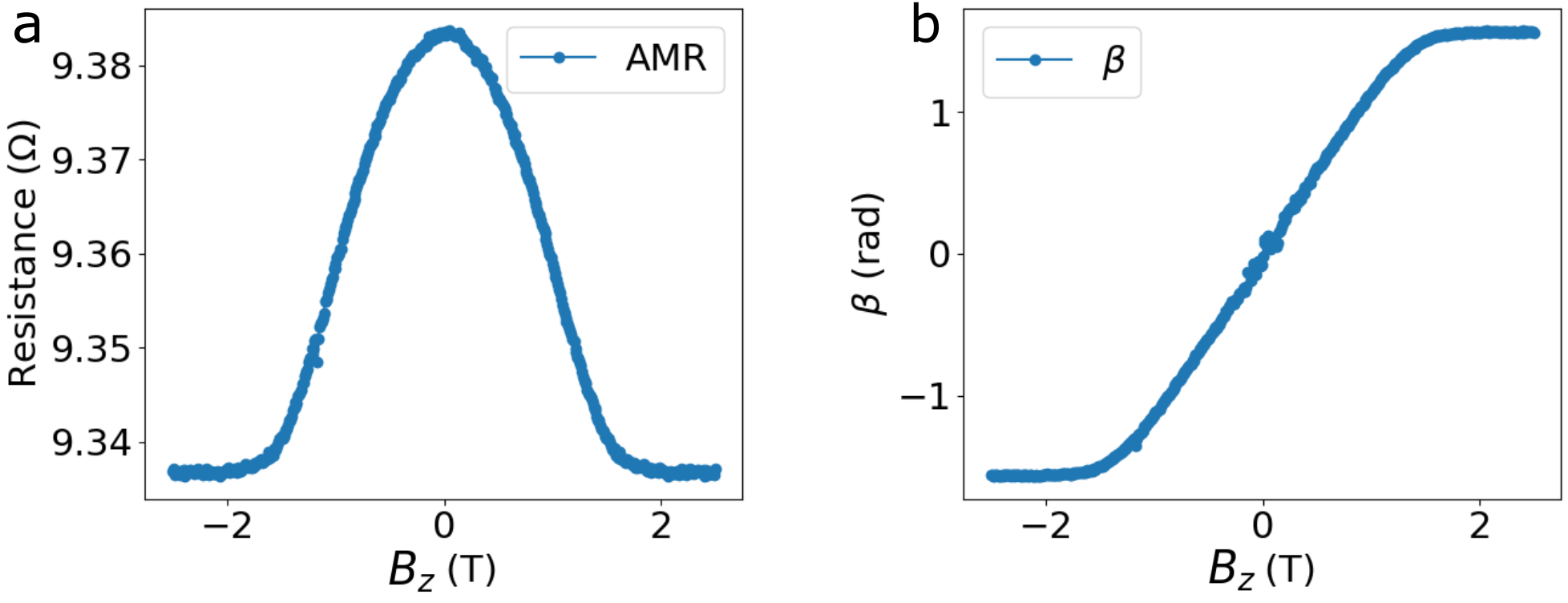}
\caption{Raw anisotropic magnetoresistance data of a Co electrode (a) and calculated values of the angle, $\beta$, between the Co magnetization and the electrode's easy axis (b).}
\label{fig:SI7}
\end{figure}

For an applied out-of-plane magnetic field, \(B_{z}\), the angle between the Co magnetization and the easy axis, \(\beta\), is determined from an anisotropic magnetoresistance (AMR) measurement of a Co element in a reference device, using the same shape and thickness of Co as in the device discussed in the main text. This could also be done using a spin diffusion measurement (as for \(B_{x}\)), although it requires a much larger field and a dedicated measurement, so for convenience, prior AMR data were used (Fig. \ref{fig:SI7}a). In this measurement, the resistance of the Co was recorded in a four-point configuration as a function of \(B_{z}\). Owing to AMR in the Co, these data reflect changes in the Co magnetization state: starting at the resistance corresponding to the Co magnetized along its easy axis and then saturating at high field. We can calculate \(\beta\) as a function of \(B_{z}\) using

\[
\rho(\beta) \;=\; \rho_{\perp} \;+\; (\rho_{\parallel} - \rho_{\perp})\,\cos^2(\beta),
\]

where \(\rho(\beta)\) is the resistivity at angle \(\beta\) between $J_\mathrm{c}$ and the magnetization. Here, \(\rho_{\perp}\) is the resistivity with the current perpendicular to the magnetization \(\bigl(B_{z} \gg 0\bigr)\), and \(\rho_{\parallel}\) is the resistivity with the current parallel to the magnetization \(\bigl(B_{z} = 0\bigr)\).

\newpage

\section{Analysis of Inhomogeneous Spin Transport Parameters}

\begin{figure}[H]
\centering
\includegraphics[width=0.8\linewidth]{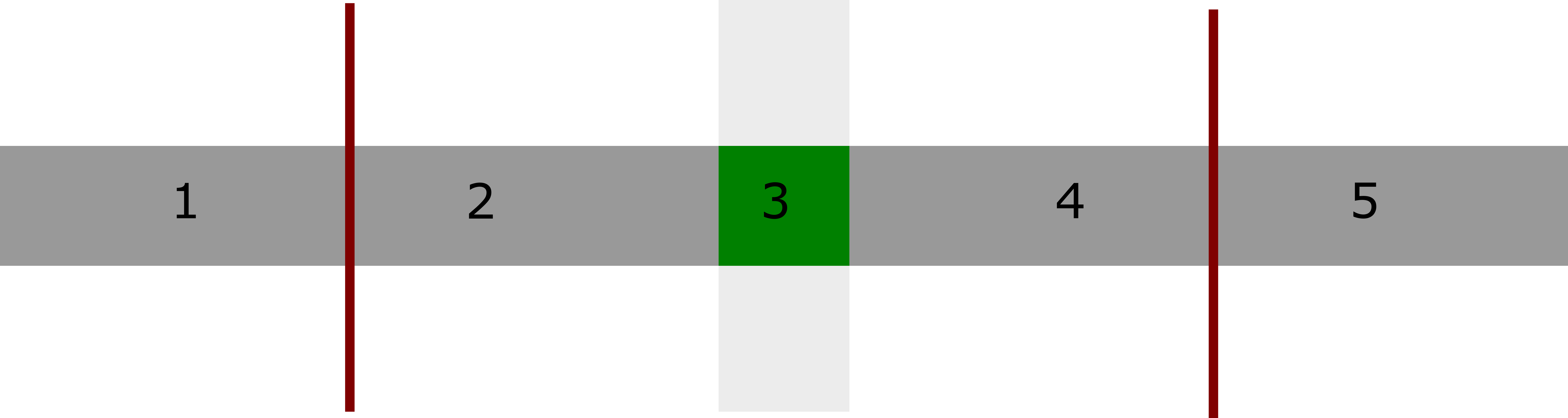}
\caption{Illustration of the 5 regions considered in this model - the side arms, used to detect voltage, are shown for illustration only, they are not considered mathematically. The FM contacts, in maroon, are likewise illustrative, and assumed to have zero width.}
\label{fig:Schematic}
\end{figure}

In the main text, we made the assumption that all spin transport parameters were the same for the pristine and the proximitized graphene. This simplification is necessary as it is not possible to experimentally probe the parameters of the two regions independently within this device architecture. To justify this assumption, we follow the method outlined in [\onlinecite{safeer_spin_2020AAA,herling_gate_2020AAA}] and employ a five-region 1D spin diffusion model. The model consists of pristine graphene to the left of the injector FM (Region 1), pristine graphene between the injector and the proximitized region (Region 2), the proximitized graphene itself (Region 3), pristine graphene between the proximitized region and the detector (Region 4), and pristine graphene after the detector (Region 5). The model is illustrated in Fig. \ref{fig:Schematic}, the dimensions of each region are taken from the device used in the work (Fig. 1, main text), with regions 1 and 5 being semi-infinite in length.

The methodology in this section uses the 5-region model as a virtual experiment. We set the "true" values of the spin transport parameters in the proximitized region ($\lambda_\mathrm{s}^\mathrm{prox}$, $\tau_\mathrm{s}^\mathrm{prox}$, and $\theta_\mathrm{SCI}$) and generate simulated Hanle and SCI datasets. These datasets are then fitted using the simpler single-region 1D spin diffusion equations to extract effective parameters ($\lambda_\mathrm{s}^\mathrm{eff}$, $\tau_\mathrm{s}^\mathrm{eff}$, and $\theta_\mathrm{SCI}^\mathrm{eff}$). This allows us to quantify the deviation of the effective parameters from the true pristine or proximitized values under different conditions. For all simulations, the parameters in the pristine graphene regions are fixed at $\lambda_\mathrm{s}^\mathrm{prist}=1000\,\mathrm{nm}$, $\tau_\mathrm{s}^\mathrm{prist}=100\,\mathrm{ps}$, $D_\mathrm{s}^\mathrm{prist}=10 \times 10^{-3}\,\mathrm{m}^2/\mathrm{s}$, and $P=0.1$.

\begin{figure}[H]
\centering
\includegraphics[width=1\linewidth]{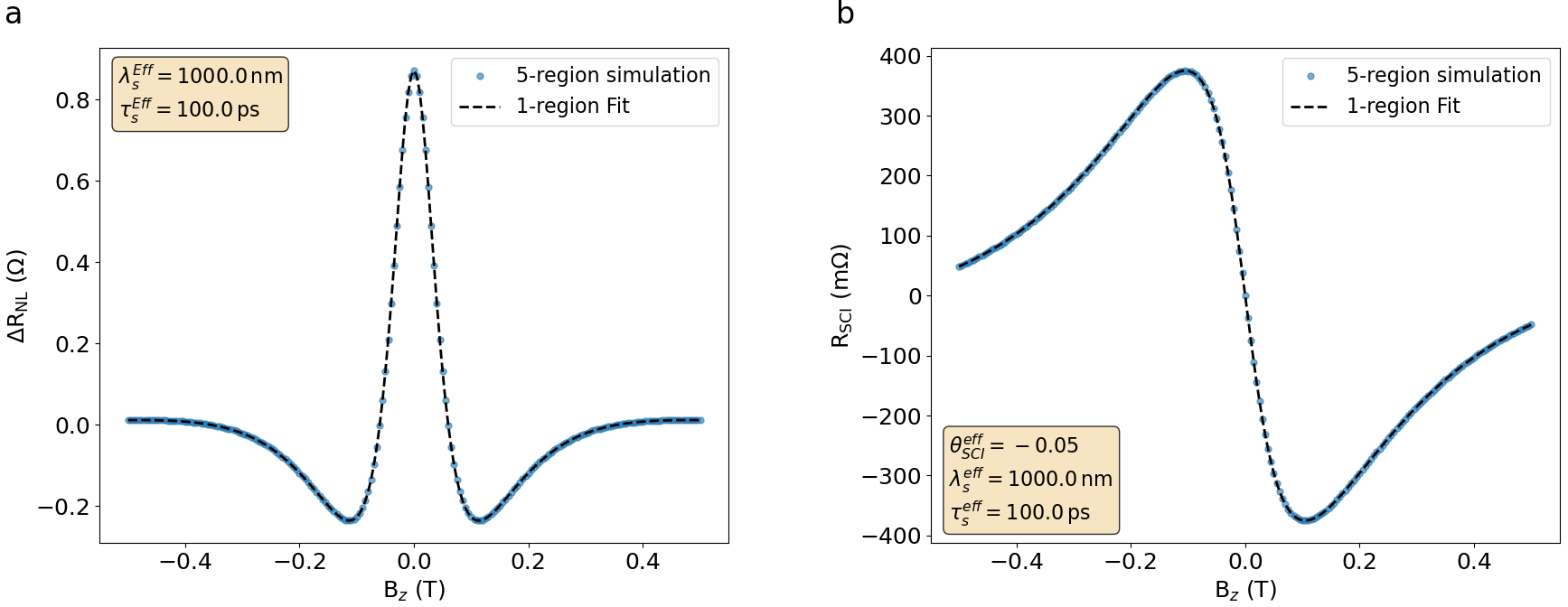}
\caption{Fitting of simulated data from the 5-region model where all regions have identical parameters. The fits use the analytical equations for (a) Hanle precession and (b) SCI. The extracted effective parameters perfectly match the true input parameters.}
\label{fig:Homogenous_1Point}
\end{figure}

As mentioned in the main text, the effective spin diffusion parameters can be extracted from either Hanle or SCI measurements. In the ideal case of a homogeneous device (i.e., where the parameters in the proximitized region are identical to the pristine region), the two fitting methods yield the same results, as shown in Fig. \ref{fig:Homogenous_1Point}.

\begin{figure}[H]
\centering
\includegraphics[width=1\linewidth]{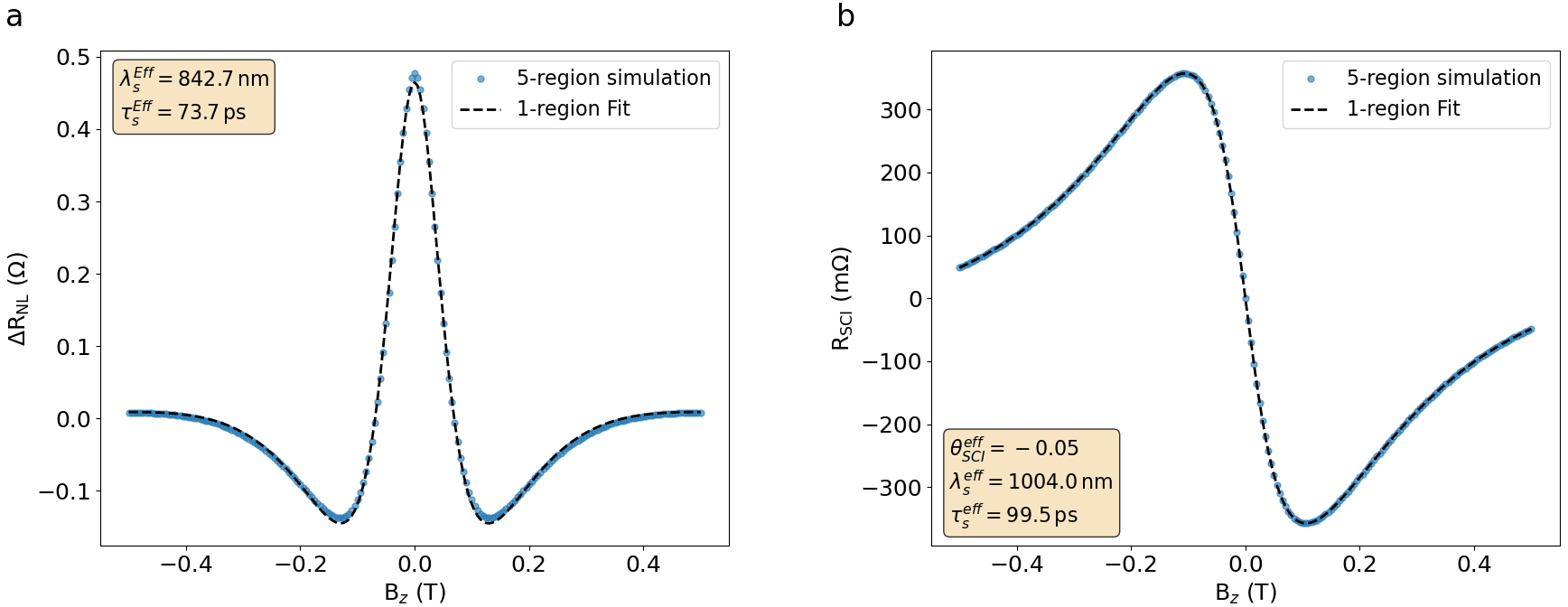}
\caption{Fitting of simulated data from the 5-region model with different spin transport parameters in the proximitized region ($\tau_\mathrm{s}^\mathrm{prox}=25$ ps, $\lambda_\mathrm{s}^\mathrm{prox}=500$ nm). The discrepancy between the Hanle and SCI fitting results is now apparent.}
\label{fig:Inhomogenous_1Point}
\end{figure}

However, when the parameters in the proximitized region are altered (e.g., $\tau_\mathrm{s}^\mathrm{prox} = 25\,\mathrm{ps}$ and $\lambda_\mathrm{s}^\mathrm{prox} = 500\,\mathrm{nm}$), while keeping the pristine values the same as before, the two fitting approaches yield different effective parameters (Fig. \ref{fig:Inhomogenous_1Point}). This demonstrates that the choice of dataset (SCI or Hanle) can influence the extracted parameters in an inhomogeneous device.

\begin{figure}[H]
\centering
\includegraphics[width=1\linewidth]{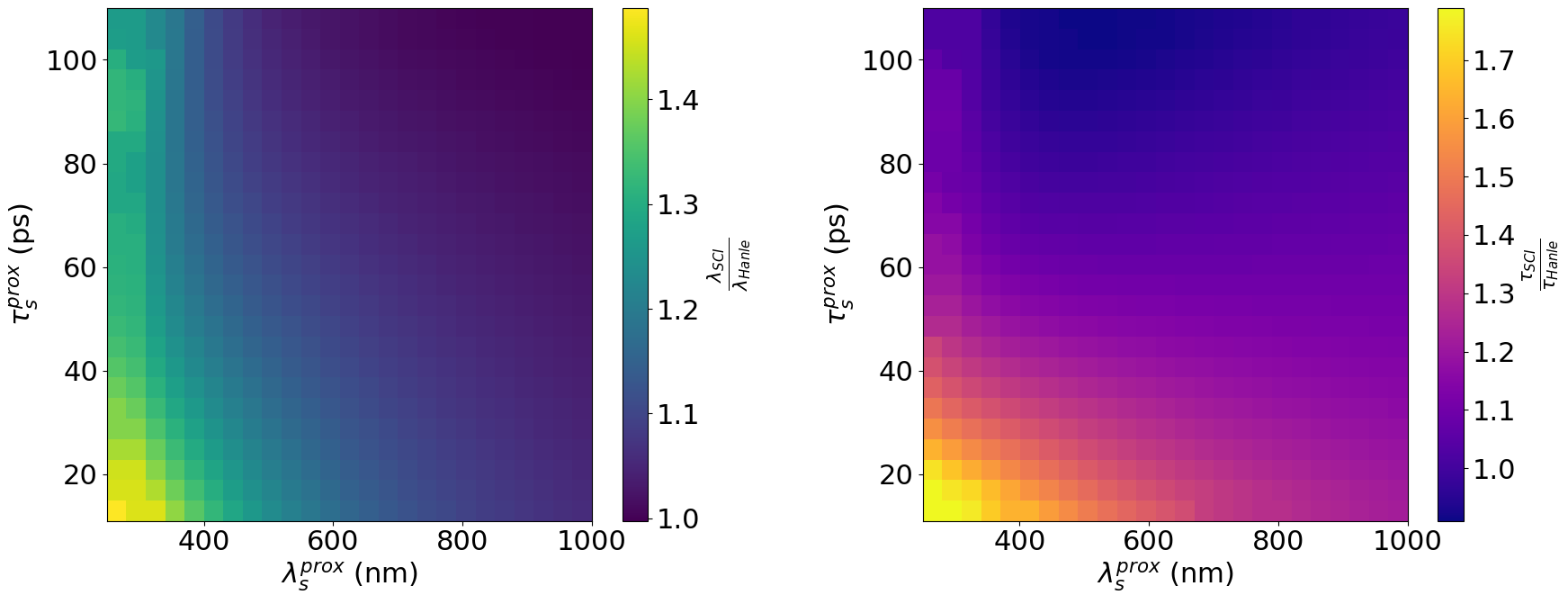}
\caption{Ratio of effective parameters extracted from SCI fitting versus Hanle fitting over a range of input $\lambda_\mathrm{s}^\mathrm{prox}$ and $\tau_\mathrm{s}^\mathrm{prox}$ values. This quantifies the discrepancy between the two fitting methods.}
\label{fig:Lambda_tau_hanle_sci}
\end{figure}

To quantify this discrepancy, we repeat the process over a grid of values for $\tau_\mathrm{s}^\mathrm{prox}$ and $\lambda_\mathrm{s}^\mathrm{prox}$, plotting the ratio of the parameters obtained from the SCI fitting to those from the Hanle fitting (Fig. \ref{fig:Lambda_tau_hanle_sci}). While the choice of method clearly makes a difference, the discrepancy between them does not exceed a factor of 2, even for extremely low values of $\tau_\mathrm{s}^\mathrm{prox}$ and $\lambda_\mathrm{s}^\mathrm{prox}$.

\begin{figure}[H]
\centering
\includegraphics[width=0.8\linewidth]{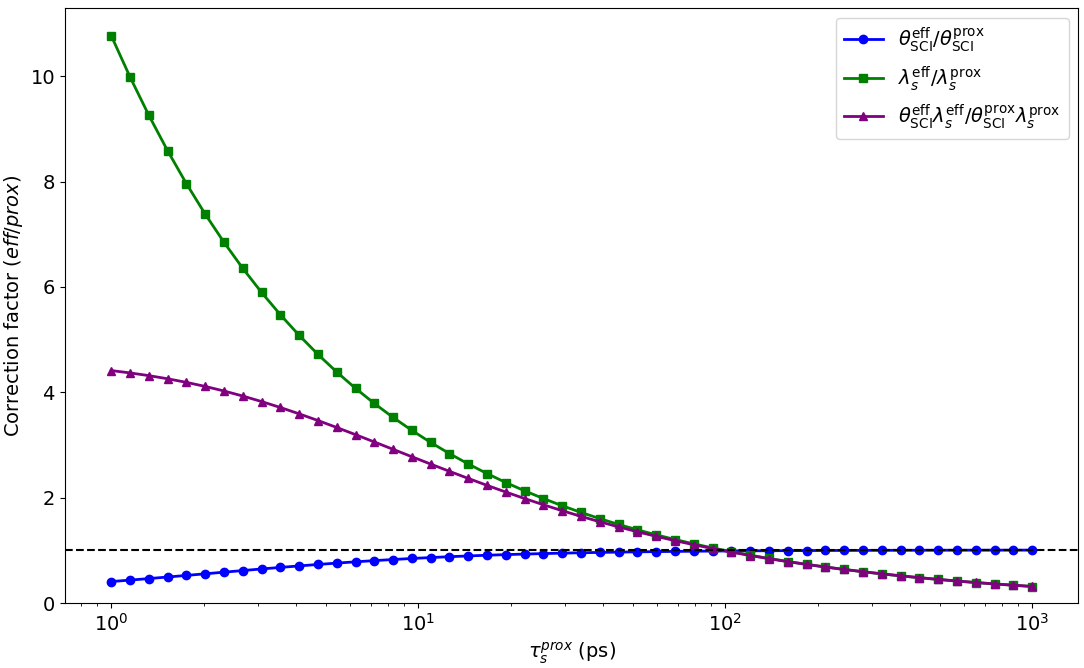}
\caption{Ratio of effective parameters (from 1D SCI fit) to true parameters (input to 5-region model) as a function of $\tau_\mathrm{s}^\mathrm{prox}$ (with $D_\mathrm{s}^\mathrm{prox}$ held constant). This shows the sensitivity of the extracted parameters to inhomogeneity in $\tau_\mathrm{s}$.} 
\label{fig:Pep_Plot}
\end{figure}

Next, we investigate how the effective parameters extracted from the SCI data deviate from the true input values - as this is the method used to extract the SCI efficiency in the main text. We use the 5-region model to produce datasets, sweeping the value of $\tau_\mathrm{s}^\mathrm{prox}$ while keeping $D_\mathrm{s}^\mathrm{prox}$ fixed (and updating $\lambda_\mathrm{s}^\mathrm{prox} = \sqrt{D_\mathrm{s}^\mathrm{prox} \, \tau_\mathrm{s}^\mathrm{prox}}$ accordingly). For each dataset, we fit the curve using the 1D model and plot the ratio of the extracted effective values to the true input values (Fig.~\ref{fig:Pep_Plot}). Interestingly, while the effective value of $\lambda_\mathrm{s}^\mathrm{eff}$ can deviate significantly from the true $\lambda_\mathrm{s}^\mathrm{prox}$, overestimating $\lambda_\mathrm{s}$ by a factor of 10 when $\lambda_\mathrm{s}^\mathrm{prox}$ is very small, the extracted $\theta_\mathrm{SCI}^\mathrm{eff}$ is much less affected, and in fact is an underestimate, with the effective value being a maximum of 2.5 times smaller than the true proximitized value. The product $\theta_\mathrm{SCI}^\mathrm{eff} \times \lambda_\mathrm{s}^\mathrm{eff}$ lies intermediate between the two individual ratios (Fig. \ref{fig:Pep_Plot}).

\begin{figure}[H]
\centering
\includegraphics[width=0.8\linewidth]{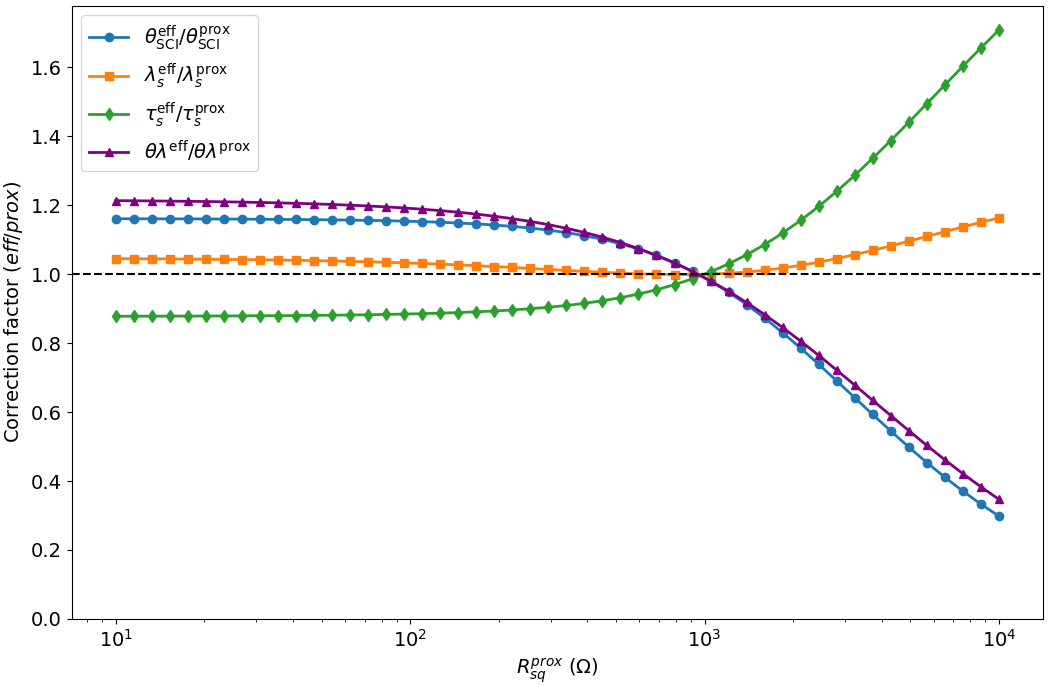}
\caption{Ratio of effective parameters to true parameters as a function of the sheet resistance in the proximitized region, $R_\mathrm{sq}^\mathrm{prox}$. The effect on the extracted parameters is minimal.}
\label{fig:Rsq_Sweep}
\end{figure}

Finally, we consider the effect of the sheet resistance, $R_\mathrm{sq}$, in the proximitized region. We perform a similar analysis, sweeping the value of $R_\mathrm{sq}^\mathrm{prox}$ while keeping other parameters fixed. The results, shown in Fig. \ref{fig:Rsq_Sweep}, indicate that variations in $R_\mathrm{sq}^\mathrm{prox}$ produce only a very minor correction to the extracted parameters over a wide range.

\newpage

\section{Fabrication Details}

The graphene/ReS$_2$ device was prepared by mechanical exfoliation, followed by viscoelastic stamping onto a clean Si/SiO$_x$ substrate, and e-beam lithography to add the metal contacts. First, graphene flakes were exfoliated and transferred onto a 300\,nm layer of SiO$_2$ on a doped Si substrate, in ambient atmosphere using blue Nitto tape. The desired flakes were identified using an optical microscope, which also allows us to determine the approximate thickness (3–4 layers) by optical contrast. This was later complemented by Raman spectroscopy (see Section S5). Next, ReS$_2$ flakes were produced by mechanical exfoliation of a bulk crystal onto polydimethylsiloxane (PDMS) (Gelpak PF GEL film WF 4, 17 mil) under an Ar atmosphere with an O$_2$ concentration below 1.0\,ppm and H$_2$O below 0.5\,ppm. The PDMS stamp, with ReS$_2$, was stamped onto the Si wafer containing the graphene using a viscoelastic stamping tool and a three-axis micrometer stage to control the position. The PDMS was removed, leaving the van der Waals heterostructure behind.

Next, the Hall bar shape was defined by thermally evaporating a hard Al mask using positive e-beam lithography (eBL). This was followed by reactive ion etching with Ar/O$_2$ plasma, removing the excess graphene. The hard mask was then removed chemically using tetra-methyl ammonium hydroxide. Next, the device was annealed at 400\,$^\circ$C for 60\,minutes in ultrahigh vacuum ($1.5 \times 10^{-8}\,\text{Torr}$) to remove fabrication residues. eBL and e-beam deposition were then used to deposit Pd/Au (5\,nm/35\,nm) contacts. Subsequently, the ferromagnetic contacts were defined by eBL, and a resistive interface was made by evaporating 0.3\,nm Ti and allowing it to oxidise in air for ten minutes. This was immediately followed by e-beam evaporation of Co/Au (35\,nm/15\,nm) and lift-off. The Au serves as a capping layer to protect the Co from oxidation.

The motivation for the Ti layer is that the spin resistance of the ferromagnet is much smaller than that of the graphene, so by placing a thin layer of $\mathrm{TiO}_x$ (0.3\,nm) we overcome the conductivity mismatch between the Co and graphene [\onlinecite{rashba_theory_2000AAA}], limiting the flow of spin current back into the ferromagnet. This leads to an interface resistance of some $\text{k}\Omega$, although with a variance between contacts depending on the exact nature of the contact. In this device, the injector $\bigl(\mathrm{F}_1\bigr)$ interface has a resistance of $6.6\,\text{k}\Omega$, as measured in a three-point configuration (voltage $\mathrm{N}_3$–$\mathrm{F}_1$, current $\mathrm{F}_1$–$\mathrm{N}_4$), while the detector interface $\bigl(\mathrm{F}_2\bigr)$ is $28.9\,\text{k}\Omega$ by the same method. The electrical measurements were performed in a Physical Property Measurement System (PPMS) by Quantum Design, using a DC reversal technique with a Keithley 2182 nanovoltmeter and a 6221 current source. The $n$-doped Si substrate acts as a back-gate electrode, to which the gate voltage ($V_\mathrm{g}$) was applied using a Keithley 2636B.


\begin{thebibliography}{44}%
\makeatletter
\providecommand \@ifxundefined [1]{%
 \@ifx{#1\undefined}
}%
\providecommand \@ifnum [1]{%
 \ifnum #1\expandafter \@firstoftwo
 \else \expandafter \@secondoftwo
 \fi
}%
\providecommand \@ifx [1]{%
 \ifx #1\expandafter \@firstoftwo
 \else \expandafter \@secondoftwo
 \fi
}%
\providecommand \natexlab [1]{#1}%
\providecommand \enquote  [1]{``#1''}%
\providecommand \bibnamefont  [1]{#1}%
\providecommand \bibfnamefont [1]{#1}%
\providecommand \citenamefont [1]{#1}%
\providecommand \href@noop [0]{\@secondoftwo}%
\providecommand \href [0]{\begingroup \@sanitize@url \@href}%
\providecommand \@href[1]{\@@startlink{#1}\@@href}%
\providecommand \@@href[1]{\endgroup#1\@@endlink}%
\providecommand \@sanitize@url [0]{\catcode `\\12\catcode `\$12\catcode `\&12\catcode `\#12\catcode `\^12\catcode `\_12\catcode `\%12\relax}%
\providecommand \@@startlink[1]{}%
\providecommand \@@endlink[0]{}%
\providecommand \url  [0]{\begingroup\@sanitize@url \@url }%
\providecommand \@url [1]{\endgroup\@href {#1}{\urlprefix }}%
\providecommand \urlprefix  [0]{URL }%
\providecommand \Eprint [0]{\href }%
\providecommand \doibase [0]{http://dx.doi.org/}%
\providecommand \selectlanguage [0]{\@gobble}%
\providecommand \bibinfo  [0]{\@secondoftwo}%
\providecommand \bibfield  [0]{\@secondoftwo}%
\providecommand \translation [1]{[#1]}%
\providecommand \BibitemOpen [0]{}%
\providecommand \bibitemStop [0]{}%
\providecommand \bibitemNoStop [0]{.\EOS\space}%
\providecommand \EOS [0]{\spacefactor3000\relax}%
\providecommand \BibitemShut  [1]{\csname bibitem#1\endcsname}%
\let\auto@bib@innerbib\@empty
\bibitem [{\citenamefont {Avsar}\ \emph {et~al.}(2020)\citenamefont {Avsar}, \citenamefont {Ochoa}, \citenamefont {Guinea}, \citenamefont {Özyilmaz}, \citenamefont {Van~Wees},\ and\ \citenamefont {Vera-Marun}}]{avsar_colloquium_2020}%
  \BibitemOpen
  \bibfield  {author} {\bibinfo {author} {\bibfnamefont {A.}~\bibnamefont {Avsar}}, \bibinfo {author} {\bibfnamefont {H.}~\bibnamefont {Ochoa}}, \bibinfo {author} {\bibfnamefont {F.}~\bibnamefont {Guinea}}, \bibinfo {author} {\bibfnamefont {B.}~\bibnamefont {Özyilmaz}}, \bibinfo {author} {\bibfnamefont {B.}~\bibnamefont {Van~Wees}}, \ and\ \bibinfo {author} {\bibfnamefont {I.}~\bibnamefont {Vera-Marun}},\ }\href {\doibase 10.1103/RevModPhys.92.021003} {\bibfield  {journal} {\bibinfo  {journal} {Reviews of Modern Physics}\ }\textbf {\bibinfo {volume} {92}},\ \bibinfo {pages} {021003} (\bibinfo {year} {2020})}\BibitemShut {NoStop}%
\bibitem [{\citenamefont {Han}\ \emph {et~al.}(2014)\citenamefont {Han}, \citenamefont {Kawakami}, \citenamefont {Gmitra},\ and\ \citenamefont {Fabian}}]{han_graphene_2014}%
  \BibitemOpen
  \bibfield  {author} {\bibinfo {author} {\bibfnamefont {W.}~\bibnamefont {Han}}, \bibinfo {author} {\bibfnamefont {R.~K.}\ \bibnamefont {Kawakami}}, \bibinfo {author} {\bibfnamefont {M.}~\bibnamefont {Gmitra}}, \ and\ \bibinfo {author} {\bibfnamefont {J.}~\bibnamefont {Fabian}},\ }\href {\doibase 10.1038/nnano.2014.214} {\bibfield  {journal} {\bibinfo  {journal} {Nature Nanotechnology}\ }\textbf {\bibinfo {volume} {9}},\ \bibinfo {pages} {794} (\bibinfo {year} {2014})}\BibitemShut {NoStop}%
\bibitem [{\citenamefont {Zhou}\ \emph {et~al.}(2024)\citenamefont {Zhou}, \citenamefont {Lu}, \citenamefont {Yang}, \citenamefont {Zhang}, \citenamefont {Liu}, \citenamefont {Zeng}, \citenamefont {Yan}, \citenamefont {Li}, \citenamefont {Wei}, \citenamefont {Wu}, \citenamefont {Pu}, \citenamefont {Liu}, \citenamefont {He}, \citenamefont {Zhang},\ and\ \citenamefont {Xu}}]{Zhou2024-tj}%
  \BibitemOpen
  \bibfield  {author} {\bibinfo {author} {\bibfnamefont {J.}~\bibnamefont {Zhou}}, \bibinfo {author} {\bibfnamefont {X.}~\bibnamefont {Lu}}, \bibinfo {author} {\bibfnamefont {J.}~\bibnamefont {Yang}}, \bibinfo {author} {\bibfnamefont {X.}~\bibnamefont {Zhang}}, \bibinfo {author} {\bibfnamefont {Q.}~\bibnamefont {Liu}}, \bibinfo {author} {\bibfnamefont {Q.}~\bibnamefont {Zeng}}, \bibinfo {author} {\bibfnamefont {Y.}~\bibnamefont {Yan}}, \bibinfo {author} {\bibfnamefont {Y.}~\bibnamefont {Li}}, \bibinfo {author} {\bibfnamefont {L.}~\bibnamefont {Wei}}, \bibinfo {author} {\bibfnamefont {J.}~\bibnamefont {Wu}}, \bibinfo {author} {\bibfnamefont {Y.}~\bibnamefont {Pu}}, \bibinfo {author} {\bibfnamefont {R.}~\bibnamefont {Liu}}, \bibinfo {author} {\bibfnamefont {L.}~\bibnamefont {He}}, \bibinfo {author} {\bibfnamefont {R.}~\bibnamefont {Zhang}}, \ and\ \bibinfo {author} {\bibfnamefont {Y.}~\bibnamefont {Xu}},\ }\href@noop {} {\bibfield  {journal} {\bibinfo  {journal} {Carbon}\ }\textbf {\bibinfo {volume} {228}},\
  \bibinfo {pages} {119321} (\bibinfo {year} {2024})}\BibitemShut {NoStop}%
\bibitem [{\citenamefont {Ingla-Aynés}\ \emph {et~al.}(2015)\citenamefont {Ingla-Aynés}, \citenamefont {Guimarães}, \citenamefont {Meijerink}, \citenamefont {Zomer},\ and\ \citenamefont {Van~Wees}}]{ingla-aynes_24_2015}%
  \BibitemOpen
  \bibfield  {author} {\bibinfo {author} {\bibfnamefont {J.}~\bibnamefont {Ingla-Aynés}}, \bibinfo {author} {\bibfnamefont {M.~H.~D.}\ \bibnamefont {Guimarães}}, \bibinfo {author} {\bibfnamefont {R.~J.}\ \bibnamefont {Meijerink}}, \bibinfo {author} {\bibfnamefont {P.~J.}\ \bibnamefont {Zomer}}, \ and\ \bibinfo {author} {\bibfnamefont {B.~J.}\ \bibnamefont {Van~Wees}},\ }\href {\doibase 10.1103/PhysRevB.92.201410} {\bibfield  {journal} {\bibinfo  {journal} {Physical Review B}\ }\textbf {\bibinfo {volume} {92}},\ \bibinfo {pages} {201410} (\bibinfo {year} {2015})}\BibitemShut {NoStop}%
\bibitem [{\citenamefont {Drögeler}\ \emph {et~al.}(2016)\citenamefont {Drögeler}, \citenamefont {Franzen}, \citenamefont {Volmer}, \citenamefont {Pohlmann}, \citenamefont {Banszerus}, \citenamefont {Wolter}, \citenamefont {Watanabe}, \citenamefont {Taniguchi}, \citenamefont {Stampfer},\ and\ \citenamefont {Beschoten}}]{drogeler_spin_2016}%
  \BibitemOpen
  \bibfield  {author} {\bibinfo {author} {\bibfnamefont {M.}~\bibnamefont {Drögeler}}, \bibinfo {author} {\bibfnamefont {C.}~\bibnamefont {Franzen}}, \bibinfo {author} {\bibfnamefont {F.}~\bibnamefont {Volmer}}, \bibinfo {author} {\bibfnamefont {T.}~\bibnamefont {Pohlmann}}, \bibinfo {author} {\bibfnamefont {L.}~\bibnamefont {Banszerus}}, \bibinfo {author} {\bibfnamefont {M.}~\bibnamefont {Wolter}}, \bibinfo {author} {\bibfnamefont {K.}~\bibnamefont {Watanabe}}, \bibinfo {author} {\bibfnamefont {T.}~\bibnamefont {Taniguchi}}, \bibinfo {author} {\bibfnamefont {C.}~\bibnamefont {Stampfer}}, \ and\ \bibinfo {author} {\bibfnamefont {B.}~\bibnamefont {Beschoten}},\ }\href {\doibase 10.1021/acs.nanolett.6b00497} {\bibfield  {journal} {\bibinfo  {journal} {Nano Letters}\ }\textbf {\bibinfo {volume} {16}},\ \bibinfo {pages} {3533} (\bibinfo {year} {2016})}\BibitemShut {NoStop}%
\bibitem [{\citenamefont {Manipatruni}\ \emph {et~al.}(2019)\citenamefont {Manipatruni}, \citenamefont {Nikonov}, \citenamefont {Lin}, \citenamefont {Gosavi}, \citenamefont {Liu}, \citenamefont {Prasad}, \citenamefont {Huang}, \citenamefont {Bonturim}, \citenamefont {Ramesh},\ and\ \citenamefont {Young}}]{manipatruni_scalable_2019}%
  \BibitemOpen
  \bibfield  {author} {\bibinfo {author} {\bibfnamefont {S.}~\bibnamefont {Manipatruni}}, \bibinfo {author} {\bibfnamefont {D.~E.}\ \bibnamefont {Nikonov}}, \bibinfo {author} {\bibfnamefont {C.-C.}\ \bibnamefont {Lin}}, \bibinfo {author} {\bibfnamefont {T.~A.}\ \bibnamefont {Gosavi}}, \bibinfo {author} {\bibfnamefont {H.}~\bibnamefont {Liu}}, \bibinfo {author} {\bibfnamefont {B.}~\bibnamefont {Prasad}}, \bibinfo {author} {\bibfnamefont {Y.-L.}\ \bibnamefont {Huang}}, \bibinfo {author} {\bibfnamefont {E.}~\bibnamefont {Bonturim}}, \bibinfo {author} {\bibfnamefont {R.}~\bibnamefont {Ramesh}}, \ and\ \bibinfo {author} {\bibfnamefont {I.~A.}\ \bibnamefont {Young}},\ }\href {\doibase 10.1038/s41586-018-0770-2} {\bibfield  {journal} {\bibinfo  {journal} {Nature}\ }\textbf {\bibinfo {volume} {565}},\ \bibinfo {pages} {35} (\bibinfo {year} {2019})}\BibitemShut {NoStop}%
\bibitem [{\citenamefont {Pham}\ \emph {et~al.}(2020)\citenamefont {Pham}, \citenamefont {Groen}, \citenamefont {Manipatruni}, \citenamefont {Choi}, \citenamefont {Nikonov}, \citenamefont {Sagasta}, \citenamefont {Lin}, \citenamefont {Gosavi}, \citenamefont {Marty}, \citenamefont {Hueso}, \citenamefont {Young},\ and\ \citenamefont {Casanova}}]{pham_spinorbit_2020}%
  \BibitemOpen
  \bibfield  {author} {\bibinfo {author} {\bibfnamefont {V.~T.}\ \bibnamefont {Pham}}, \bibinfo {author} {\bibfnamefont {I.}~\bibnamefont {Groen}}, \bibinfo {author} {\bibfnamefont {S.}~\bibnamefont {Manipatruni}}, \bibinfo {author} {\bibfnamefont {W.~Y.}\ \bibnamefont {Choi}}, \bibinfo {author} {\bibfnamefont {D.~E.}\ \bibnamefont {Nikonov}}, \bibinfo {author} {\bibfnamefont {E.}~\bibnamefont {Sagasta}}, \bibinfo {author} {\bibfnamefont {C.-C.}\ \bibnamefont {Lin}}, \bibinfo {author} {\bibfnamefont {T.~A.}\ \bibnamefont {Gosavi}}, \bibinfo {author} {\bibfnamefont {A.}~\bibnamefont {Marty}}, \bibinfo {author} {\bibfnamefont {L.~E.}\ \bibnamefont {Hueso}}, \bibinfo {author} {\bibfnamefont {I.~A.}\ \bibnamefont {Young}}, \ and\ \bibinfo {author} {\bibfnamefont {F.}~\bibnamefont {Casanova}},\ }\href {\doibase 10.1038/s41928-020-0395-y} {\bibfield  {journal} {\bibinfo  {journal} {Nature Electronics}\ }\textbf {\bibinfo {volume} {3}},\ \bibinfo {pages} {309} (\bibinfo {year} {2020})}\BibitemShut {NoStop}%
\bibitem [{\citenamefont {Vaz}\ \emph {et~al.}(2024)\citenamefont {Vaz}, \citenamefont {Lin}, \citenamefont {Plombon}, \citenamefont {Choi}, \citenamefont {Groen}, \citenamefont {Arango}, \citenamefont {Chuvilin}, \citenamefont {Hueso}, \citenamefont {Nikonov}, \citenamefont {Li}, \citenamefont {Debashis}, \citenamefont {Clendenning}, \citenamefont {Gosavi}, \citenamefont {Huang}, \citenamefont {Prasad}, \citenamefont {Ramesh}, \citenamefont {Vecchiola}, \citenamefont {Bibes}, \citenamefont {Bouzehouane}, \citenamefont {Fusil}, \citenamefont {Garcia}, \citenamefont {Young},\ and\ \citenamefont {Casanova}}]{vaz_voltage-based_2024}%
  \BibitemOpen
  \bibfield  {author} {\bibinfo {author} {\bibfnamefont {D.~C.}\ \bibnamefont {Vaz}}, \bibinfo {author} {\bibfnamefont {C.-C.}\ \bibnamefont {Lin}}, \bibinfo {author} {\bibfnamefont {J.~J.}\ \bibnamefont {Plombon}}, \bibinfo {author} {\bibfnamefont {W.~Y.}\ \bibnamefont {Choi}}, \bibinfo {author} {\bibfnamefont {I.}~\bibnamefont {Groen}}, \bibinfo {author} {\bibfnamefont {I.~C.}\ \bibnamefont {Arango}}, \bibinfo {author} {\bibfnamefont {A.}~\bibnamefont {Chuvilin}}, \bibinfo {author} {\bibfnamefont {L.~E.}\ \bibnamefont {Hueso}}, \bibinfo {author} {\bibfnamefont {D.~E.}\ \bibnamefont {Nikonov}}, \bibinfo {author} {\bibfnamefont {H.}~\bibnamefont {Li}}, \bibinfo {author} {\bibfnamefont {P.}~\bibnamefont {Debashis}}, \bibinfo {author} {\bibfnamefont {S.~B.}\ \bibnamefont {Clendenning}}, \bibinfo {author} {\bibfnamefont {T.~A.}\ \bibnamefont {Gosavi}}, \bibinfo {author} {\bibfnamefont {Y.-L.}\ \bibnamefont {Huang}}, \bibinfo {author} {\bibfnamefont {B.}~\bibnamefont {Prasad}}, \bibinfo {author} {\bibfnamefont
  {R.}~\bibnamefont {Ramesh}}, \bibinfo {author} {\bibfnamefont {A.}~\bibnamefont {Vecchiola}}, \bibinfo {author} {\bibfnamefont {M.}~\bibnamefont {Bibes}}, \bibinfo {author} {\bibfnamefont {K.}~\bibnamefont {Bouzehouane}}, \bibinfo {author} {\bibfnamefont {S.}~\bibnamefont {Fusil}}, \bibinfo {author} {\bibfnamefont {V.}~\bibnamefont {Garcia}}, \bibinfo {author} {\bibfnamefont {I.~A.}\ \bibnamefont {Young}}, \ and\ \bibinfo {author} {\bibfnamefont {F.}~\bibnamefont {Casanova}},\ }\href {\doibase 10.1038/s41467-024-45868-x} {\bibfield  {journal} {\bibinfo  {journal} {Nature Communications}\ }\textbf {\bibinfo {volume} {15}},\ \bibinfo {pages} {1902} (\bibinfo {year} {2024})}\BibitemShut {NoStop}%
\bibitem [{\citenamefont {Noël}\ \emph {et~al.}(2020)\citenamefont {Noël}, \citenamefont {Trier}, \citenamefont {Vicente~Arche}, \citenamefont {Bréhin}, \citenamefont {Vaz}, \citenamefont {Garcia}, \citenamefont {Fusil}, \citenamefont {Barthélémy}, \citenamefont {Vila}, \citenamefont {Bibes},\ and\ \citenamefont {Attané}}]{noel_non-volatile_2020}%
  \BibitemOpen
  \bibfield  {author} {\bibinfo {author} {\bibfnamefont {P.}~\bibnamefont {Noël}}, \bibinfo {author} {\bibfnamefont {F.}~\bibnamefont {Trier}}, \bibinfo {author} {\bibfnamefont {L.~M.}\ \bibnamefont {Vicente~Arche}}, \bibinfo {author} {\bibfnamefont {J.}~\bibnamefont {Bréhin}}, \bibinfo {author} {\bibfnamefont {D.~C.}\ \bibnamefont {Vaz}}, \bibinfo {author} {\bibfnamefont {V.}~\bibnamefont {Garcia}}, \bibinfo {author} {\bibfnamefont {S.}~\bibnamefont {Fusil}}, \bibinfo {author} {\bibfnamefont {A.}~\bibnamefont {Barthélémy}}, \bibinfo {author} {\bibfnamefont {L.}~\bibnamefont {Vila}}, \bibinfo {author} {\bibfnamefont {M.}~\bibnamefont {Bibes}}, \ and\ \bibinfo {author} {\bibfnamefont {J.-P.}\ \bibnamefont {Attané}},\ }\href {\doibase 10.1038/s41586-020-2197-9} {\bibfield  {journal} {\bibinfo  {journal} {Nature}\ }\textbf {\bibinfo {volume} {580}},\ \bibinfo {pages} {483} (\bibinfo {year} {2020})}\BibitemShut {NoStop}%
\bibitem [{\citenamefont {Varotto}\ \emph {et~al.}(2021)\citenamefont {Varotto}, \citenamefont {Nessi}, \citenamefont {Cecchi}, \citenamefont {S{\l}awi{\'n}ska}, \citenamefont {No{\"e}l}, \citenamefont {Petr{\`o}}, \citenamefont {Fagiani}, \citenamefont {Novati}, \citenamefont {Cantoni}, \citenamefont {Petti}, \citenamefont {Albisetti}, \citenamefont {Costa}, \citenamefont {Calarco}, \citenamefont {Buongiorno~Nardelli}, \citenamefont {Bibes}, \citenamefont {Picozzi}, \citenamefont {Attan{\'e}}, \citenamefont {Vila}, \citenamefont {Bertacco},\ and\ \citenamefont {Rinaldi}}]{Varotto2021-fo}%
  \BibitemOpen
  \bibfield  {author} {\bibinfo {author} {\bibfnamefont {S.}~\bibnamefont {Varotto}}, \bibinfo {author} {\bibfnamefont {L.}~\bibnamefont {Nessi}}, \bibinfo {author} {\bibfnamefont {S.}~\bibnamefont {Cecchi}}, \bibinfo {author} {\bibfnamefont {J.}~\bibnamefont {S{\l}awi{\'n}ska}}, \bibinfo {author} {\bibfnamefont {P.}~\bibnamefont {No{\"e}l}}, \bibinfo {author} {\bibfnamefont {S.}~\bibnamefont {Petr{\`o}}}, \bibinfo {author} {\bibfnamefont {F.}~\bibnamefont {Fagiani}}, \bibinfo {author} {\bibfnamefont {A.}~\bibnamefont {Novati}}, \bibinfo {author} {\bibfnamefont {M.}~\bibnamefont {Cantoni}}, \bibinfo {author} {\bibfnamefont {D.}~\bibnamefont {Petti}}, \bibinfo {author} {\bibfnamefont {E.}~\bibnamefont {Albisetti}}, \bibinfo {author} {\bibfnamefont {M.}~\bibnamefont {Costa}}, \bibinfo {author} {\bibfnamefont {R.}~\bibnamefont {Calarco}}, \bibinfo {author} {\bibfnamefont {M.}~\bibnamefont {Buongiorno~Nardelli}}, \bibinfo {author} {\bibfnamefont {M.}~\bibnamefont {Bibes}}, \bibinfo {author} {\bibfnamefont
  {S.}~\bibnamefont {Picozzi}}, \bibinfo {author} {\bibfnamefont {J.-P.}\ \bibnamefont {Attan{\'e}}}, \bibinfo {author} {\bibfnamefont {L.}~\bibnamefont {Vila}}, \bibinfo {author} {\bibfnamefont {R.}~\bibnamefont {Bertacco}}, \ and\ \bibinfo {author} {\bibfnamefont {C.}~\bibnamefont {Rinaldi}},\ }\href@noop {} {\bibfield  {journal} {\bibinfo  {journal} {Nat. Electron.}\ }\textbf {\bibinfo {volume} {4}},\ \bibinfo {pages} {740} (\bibinfo {year} {2021})}\BibitemShut {NoStop}%
\bibitem [{\citenamefont {Yang}\ \emph {et~al.}(2022)\citenamefont {Yang}, \citenamefont {Valenzuela}, \citenamefont {Chshiev}, \citenamefont {Couet}, \citenamefont {Dieny}, \citenamefont {Dlubak}, \citenamefont {Fert}, \citenamefont {Garello}, \citenamefont {Jamet}, \citenamefont {Jeong}, \citenamefont {Lee}, \citenamefont {Lee}, \citenamefont {Martin}, \citenamefont {Kar}, \citenamefont {Sénéor}, \citenamefont {Shin},\ and\ \citenamefont {Roche}}]{yang_two-dimensional_2022}%
  \BibitemOpen
  \bibfield  {author} {\bibinfo {author} {\bibfnamefont {H.}~\bibnamefont {Yang}}, \bibinfo {author} {\bibfnamefont {S.~O.}\ \bibnamefont {Valenzuela}}, \bibinfo {author} {\bibfnamefont {M.}~\bibnamefont {Chshiev}}, \bibinfo {author} {\bibfnamefont {S.}~\bibnamefont {Couet}}, \bibinfo {author} {\bibfnamefont {B.}~\bibnamefont {Dieny}}, \bibinfo {author} {\bibfnamefont {B.}~\bibnamefont {Dlubak}}, \bibinfo {author} {\bibfnamefont {A.}~\bibnamefont {Fert}}, \bibinfo {author} {\bibfnamefont {K.}~\bibnamefont {Garello}}, \bibinfo {author} {\bibfnamefont {M.}~\bibnamefont {Jamet}}, \bibinfo {author} {\bibfnamefont {D.-E.}\ \bibnamefont {Jeong}}, \bibinfo {author} {\bibfnamefont {K.}~\bibnamefont {Lee}}, \bibinfo {author} {\bibfnamefont {T.}~\bibnamefont {Lee}}, \bibinfo {author} {\bibfnamefont {M.-B.}\ \bibnamefont {Martin}}, \bibinfo {author} {\bibfnamefont {G.~S.}\ \bibnamefont {Kar}}, \bibinfo {author} {\bibfnamefont {P.}~\bibnamefont {Sénéor}}, \bibinfo {author} {\bibfnamefont {H.-J.}\ \bibnamefont {Shin}},
  \ and\ \bibinfo {author} {\bibfnamefont {S.}~\bibnamefont {Roche}},\ }\href {\doibase 10.1038/s41586-022-04768-0} {\bibfield  {journal} {\bibinfo  {journal} {Nature}\ }\textbf {\bibinfo {volume} {606}},\ \bibinfo {pages} {663} (\bibinfo {year} {2022})}\BibitemShut {NoStop}%
\bibitem [{\citenamefont {Lin}\ \emph {et~al.}(2013)\citenamefont {Lin}, \citenamefont {Penumatcha}, \citenamefont {Gao}, \citenamefont {Diep}, \citenamefont {Appenzeller},\ and\ \citenamefont {Chen}}]{lin_spin_2013}%
  \BibitemOpen
  \bibfield  {author} {\bibinfo {author} {\bibfnamefont {C.-C.}\ \bibnamefont {Lin}}, \bibinfo {author} {\bibfnamefont {A.~V.}\ \bibnamefont {Penumatcha}}, \bibinfo {author} {\bibfnamefont {Y.}~\bibnamefont {Gao}}, \bibinfo {author} {\bibfnamefont {V.~Q.}\ \bibnamefont {Diep}}, \bibinfo {author} {\bibfnamefont {J.}~\bibnamefont {Appenzeller}}, \ and\ \bibinfo {author} {\bibfnamefont {Z.}~\bibnamefont {Chen}},\ }\href {\doibase 10.1021/nl402547m} {\bibfield  {journal} {\bibinfo  {journal} {Nano Letters}\ }\textbf {\bibinfo {volume} {13}},\ \bibinfo {pages} {5177} (\bibinfo {year} {2013})}\BibitemShut {NoStop}%
\bibitem [{\citenamefont {Weeks}\ \emph {et~al.}(2011)\citenamefont {Weeks}, \citenamefont {Hu}, \citenamefont {Alicea}, \citenamefont {Franz},\ and\ \citenamefont {Wu}}]{weeks_engineering_2011}%
  \BibitemOpen
  \bibfield  {author} {\bibinfo {author} {\bibfnamefont {C.}~\bibnamefont {Weeks}}, \bibinfo {author} {\bibfnamefont {J.}~\bibnamefont {Hu}}, \bibinfo {author} {\bibfnamefont {J.}~\bibnamefont {Alicea}}, \bibinfo {author} {\bibfnamefont {M.}~\bibnamefont {Franz}}, \ and\ \bibinfo {author} {\bibfnamefont {R.}~\bibnamefont {Wu}},\ }\href {\doibase 10.1103/PhysRevX.1.021001} {\bibfield  {journal} {\bibinfo  {journal} {Physical Review X}\ }\textbf {\bibinfo {volume} {1}},\ \bibinfo {pages} {021001} (\bibinfo {year} {2011})}\BibitemShut {NoStop}%
\bibitem [{\citenamefont {Gmitra}, \citenamefont {Kochan},\ and\ \citenamefont {Fabian}(2013)}]{gmitra_spin-orbit_2013}%
  \BibitemOpen
  \bibfield  {author} {\bibinfo {author} {\bibfnamefont {M.}~\bibnamefont {Gmitra}}, \bibinfo {author} {\bibfnamefont {D.}~\bibnamefont {Kochan}}, \ and\ \bibinfo {author} {\bibfnamefont {J.}~\bibnamefont {Fabian}},\ }\href {\doibase 10.1103/PhysRevLett.110.246602} {\bibfield  {journal} {\bibinfo  {journal} {Physical Review Letters}\ }\textbf {\bibinfo {volume} {110}},\ \bibinfo {pages} {246602} (\bibinfo {year} {2013})}\BibitemShut {NoStop}%
\bibitem [{\citenamefont {Calleja}\ \emph {et~al.}(2015)\citenamefont {Calleja}, \citenamefont {Ochoa}, \citenamefont {Garnica}, \citenamefont {Barja}, \citenamefont {Navarro}, \citenamefont {Black}, \citenamefont {Otrokov}, \citenamefont {Chulkov}, \citenamefont {Arnau}, \citenamefont {Vázquez De~Parga}, \citenamefont {Guinea},\ and\ \citenamefont {Miranda}}]{calleja_spatial_2015}%
  \BibitemOpen
  \bibfield  {author} {\bibinfo {author} {\bibfnamefont {F.}~\bibnamefont {Calleja}}, \bibinfo {author} {\bibfnamefont {H.}~\bibnamefont {Ochoa}}, \bibinfo {author} {\bibfnamefont {M.}~\bibnamefont {Garnica}}, \bibinfo {author} {\bibfnamefont {S.}~\bibnamefont {Barja}}, \bibinfo {author} {\bibfnamefont {J.~J.}\ \bibnamefont {Navarro}}, \bibinfo {author} {\bibfnamefont {A.}~\bibnamefont {Black}}, \bibinfo {author} {\bibfnamefont {M.~M.}\ \bibnamefont {Otrokov}}, \bibinfo {author} {\bibfnamefont {E.~V.}\ \bibnamefont {Chulkov}}, \bibinfo {author} {\bibfnamefont {A.}~\bibnamefont {Arnau}}, \bibinfo {author} {\bibfnamefont {A.~L.}\ \bibnamefont {Vázquez De~Parga}}, \bibinfo {author} {\bibfnamefont {F.}~\bibnamefont {Guinea}}, \ and\ \bibinfo {author} {\bibfnamefont {R.}~\bibnamefont {Miranda}},\ }\href {\doibase 10.1038/nphys3173} {\bibfield  {journal} {\bibinfo  {journal} {Nature Physics}\ }\textbf {\bibinfo {volume} {11}},\ \bibinfo {pages} {43} (\bibinfo {year} {2015})}\BibitemShut {NoStop}%
\bibitem [{\citenamefont {Gmitra}\ \emph {et~al.}(2016)\citenamefont {Gmitra}, \citenamefont {Kochan}, \citenamefont {Högl},\ and\ \citenamefont {Fabian}}]{gmitra_trivial_2016}%
  \BibitemOpen
  \bibfield  {author} {\bibinfo {author} {\bibfnamefont {M.}~\bibnamefont {Gmitra}}, \bibinfo {author} {\bibfnamefont {D.}~\bibnamefont {Kochan}}, \bibinfo {author} {\bibfnamefont {P.}~\bibnamefont {Högl}}, \ and\ \bibinfo {author} {\bibfnamefont {J.}~\bibnamefont {Fabian}},\ }\href {\doibase 10.1103/PhysRevB.93.155104} {\bibfield  {journal} {\bibinfo  {journal} {Physical Review B}\ }\textbf {\bibinfo {volume} {93}},\ \bibinfo {pages} {155104} (\bibinfo {year} {2016})}\BibitemShut {NoStop}%
\bibitem [{\citenamefont {Safeer}\ \emph {et~al.}(2019{\natexlab{a}})\citenamefont {Safeer}, \citenamefont {Ingla-Aynés}, \citenamefont {Herling}, \citenamefont {Garcia}, \citenamefont {Vila}, \citenamefont {Ontoso}, \citenamefont {Calvo}, \citenamefont {Roche}, \citenamefont {Hueso},\ and\ \citenamefont {Casanova}}]{safeer_room-temperature_2019}%
  \BibitemOpen
  \bibfield  {author} {\bibinfo {author} {\bibfnamefont {C.~K.}\ \bibnamefont {Safeer}}, \bibinfo {author} {\bibfnamefont {J.}~\bibnamefont {Ingla-Aynés}}, \bibinfo {author} {\bibfnamefont {F.}~\bibnamefont {Herling}}, \bibinfo {author} {\bibfnamefont {J.~H.}\ \bibnamefont {Garcia}}, \bibinfo {author} {\bibfnamefont {M.}~\bibnamefont {Vila}}, \bibinfo {author} {\bibfnamefont {N.}~\bibnamefont {Ontoso}}, \bibinfo {author} {\bibfnamefont {M.~R.}\ \bibnamefont {Calvo}}, \bibinfo {author} {\bibfnamefont {S.}~\bibnamefont {Roche}}, \bibinfo {author} {\bibfnamefont {L.~E.}\ \bibnamefont {Hueso}}, \ and\ \bibinfo {author} {\bibfnamefont {F.}~\bibnamefont {Casanova}},\ }\href {\doibase 10.1021/acs.nanolett.8b04368} {\bibfield  {journal} {\bibinfo  {journal} {Nano Letters}\ }\textbf {\bibinfo {volume} {19}},\ \bibinfo {pages} {1074} (\bibinfo {year} {2019}{\natexlab{a}})}\BibitemShut {NoStop}%
\bibitem [{\citenamefont {Safeer}\ \emph {et~al.}(2019{\natexlab{b}})\citenamefont {Safeer}, \citenamefont {Ontoso}, \citenamefont {Ingla-Aynés}, \citenamefont {Herling}, \citenamefont {Pham}, \citenamefont {Kurzmann}, \citenamefont {Ensslin}, \citenamefont {Chuvilin}, \citenamefont {Robredo}, \citenamefont {Vergniory}, \citenamefont {De~Juan}, \citenamefont {Hueso}, \citenamefont {Calvo},\ and\ \citenamefont {Casanova}}]{safeer_large_2019}%
  \BibitemOpen
  \bibfield  {author} {\bibinfo {author} {\bibfnamefont {C.~K.}\ \bibnamefont {Safeer}}, \bibinfo {author} {\bibfnamefont {N.}~\bibnamefont {Ontoso}}, \bibinfo {author} {\bibfnamefont {J.}~\bibnamefont {Ingla-Aynés}}, \bibinfo {author} {\bibfnamefont {F.}~\bibnamefont {Herling}}, \bibinfo {author} {\bibfnamefont {V.~T.}\ \bibnamefont {Pham}}, \bibinfo {author} {\bibfnamefont {A.}~\bibnamefont {Kurzmann}}, \bibinfo {author} {\bibfnamefont {K.}~\bibnamefont {Ensslin}}, \bibinfo {author} {\bibfnamefont {A.}~\bibnamefont {Chuvilin}}, \bibinfo {author} {\bibfnamefont {I.}~\bibnamefont {Robredo}}, \bibinfo {author} {\bibfnamefont {M.~G.}\ \bibnamefont {Vergniory}}, \bibinfo {author} {\bibfnamefont {F.}~\bibnamefont {De~Juan}}, \bibinfo {author} {\bibfnamefont {L.~E.}\ \bibnamefont {Hueso}}, \bibinfo {author} {\bibfnamefont {M.~R.}\ \bibnamefont {Calvo}}, \ and\ \bibinfo {author} {\bibfnamefont {F.}~\bibnamefont {Casanova}},\ }\href {\doibase 10.1021/acs.nanolett.9b03485} {\bibfield  {journal} {\bibinfo  {journal}
  {Nano Letters}\ }\textbf {\bibinfo {volume} {19}},\ \bibinfo {pages} {8758} (\bibinfo {year} {2019}{\natexlab{b}})}\BibitemShut {NoStop}%
\bibitem [{\citenamefont {Ghiasi}\ \emph {et~al.}(2019)\citenamefont {Ghiasi}, \citenamefont {Kaverzin}, \citenamefont {Blah},\ and\ \citenamefont {Van~Wees}}]{ghiasi_charge--spin_2019}%
  \BibitemOpen
  \bibfield  {author} {\bibinfo {author} {\bibfnamefont {T.~S.}\ \bibnamefont {Ghiasi}}, \bibinfo {author} {\bibfnamefont {A.~A.}\ \bibnamefont {Kaverzin}}, \bibinfo {author} {\bibfnamefont {P.~J.}\ \bibnamefont {Blah}}, \ and\ \bibinfo {author} {\bibfnamefont {B.~J.}\ \bibnamefont {Van~Wees}},\ }\href {\doibase 10.1021/acs.nanolett.9b01611} {\bibfield  {journal} {\bibinfo  {journal} {Nano Letters}\ }\textbf {\bibinfo {volume} {19}},\ \bibinfo {pages} {5959} (\bibinfo {year} {2019})}\BibitemShut {NoStop}%
\bibitem [{\citenamefont {Khokhriakov}\ \emph {et~al.}(2020)\citenamefont {Khokhriakov}, \citenamefont {Hoque}, \citenamefont {Karpiak},\ and\ \citenamefont {Dash}}]{khokhriakov_gate-tunable_2020}%
  \BibitemOpen
  \bibfield  {author} {\bibinfo {author} {\bibfnamefont {D.}~\bibnamefont {Khokhriakov}}, \bibinfo {author} {\bibfnamefont {A.~M.}\ \bibnamefont {Hoque}}, \bibinfo {author} {\bibfnamefont {B.}~\bibnamefont {Karpiak}}, \ and\ \bibinfo {author} {\bibfnamefont {S.~P.}\ \bibnamefont {Dash}},\ }\href {\doibase 10.1038/s41467-020-17481-1} {\bibfield  {journal} {\bibinfo  {journal} {Nature Communications}\ }\textbf {\bibinfo {volume} {11}},\ \bibinfo {pages} {3657} (\bibinfo {year} {2020})}\BibitemShut {NoStop}%
\bibitem [{\citenamefont {Herling}\ \emph {et~al.}(2020)\citenamefont {Herling}, \citenamefont {Safeer}, \citenamefont {Ingla-Aynés}, \citenamefont {Ontoso}, \citenamefont {Hueso},\ and\ \citenamefont {Casanova}}]{herling_gate_2020}%
  \BibitemOpen
  \bibfield  {author} {\bibinfo {author} {\bibfnamefont {F.}~\bibnamefont {Herling}}, \bibinfo {author} {\bibfnamefont {C.~K.}\ \bibnamefont {Safeer}}, \bibinfo {author} {\bibfnamefont {J.}~\bibnamefont {Ingla-Aynés}}, \bibinfo {author} {\bibfnamefont {N.}~\bibnamefont {Ontoso}}, \bibinfo {author} {\bibfnamefont {L.~E.}\ \bibnamefont {Hueso}}, \ and\ \bibinfo {author} {\bibfnamefont {F.}~\bibnamefont {Casanova}},\ }\href {\doibase 10.1063/5.0006101} {\bibfield  {journal} {\bibinfo  {journal} {APL Materials}\ }\textbf {\bibinfo {volume} {8}},\ \bibinfo {pages} {071103} (\bibinfo {year} {2020})}\BibitemShut {NoStop}%
\bibitem [{\citenamefont {Safeer}\ \emph {et~al.}(2020)\citenamefont {Safeer}, \citenamefont {Ingla-Aynés}, \citenamefont {Ontoso}, \citenamefont {Herling}, \citenamefont {Yan}, \citenamefont {Hueso},\ and\ \citenamefont {Casanova}}]{safeer_spin_2020}%
  \BibitemOpen
  \bibfield  {author} {\bibinfo {author} {\bibfnamefont {C.~K.}\ \bibnamefont {Safeer}}, \bibinfo {author} {\bibfnamefont {J.}~\bibnamefont {Ingla-Aynés}}, \bibinfo {author} {\bibfnamefont {N.}~\bibnamefont {Ontoso}}, \bibinfo {author} {\bibfnamefont {F.}~\bibnamefont {Herling}}, \bibinfo {author} {\bibfnamefont {W.}~\bibnamefont {Yan}}, \bibinfo {author} {\bibfnamefont {L.~E.}\ \bibnamefont {Hueso}}, \ and\ \bibinfo {author} {\bibfnamefont {F.}~\bibnamefont {Casanova}},\ }\href {\doibase 10.1021/acs.nanolett.0c01428} {\bibfield  {journal} {\bibinfo  {journal} {Nano Letters}\ }\textbf {\bibinfo {volume} {20}},\ \bibinfo {pages} {4573} (\bibinfo {year} {2020})}\BibitemShut {NoStop}%
\bibitem [{\citenamefont {Benítez}\ \emph {et~al.}(2020)\citenamefont {Benítez}, \citenamefont {Savero~Torres}, \citenamefont {Sierra}, \citenamefont {Timmermans}, \citenamefont {Garcia}, \citenamefont {Roche}, \citenamefont {Costache},\ and\ \citenamefont {Valenzuela}}]{benitez_tunable_2020}%
  \BibitemOpen
  \bibfield  {author} {\bibinfo {author} {\bibfnamefont {L.~A.}\ \bibnamefont {Benítez}}, \bibinfo {author} {\bibfnamefont {W.}~\bibnamefont {Savero~Torres}}, \bibinfo {author} {\bibfnamefont {J.~F.}\ \bibnamefont {Sierra}}, \bibinfo {author} {\bibfnamefont {M.}~\bibnamefont {Timmermans}}, \bibinfo {author} {\bibfnamefont {J.~H.}\ \bibnamefont {Garcia}}, \bibinfo {author} {\bibfnamefont {S.}~\bibnamefont {Roche}}, \bibinfo {author} {\bibfnamefont {M.~V.}\ \bibnamefont {Costache}}, \ and\ \bibinfo {author} {\bibfnamefont {S.~O.}\ \bibnamefont {Valenzuela}},\ }\href {\doibase 10.1038/s41563-019-0575-1} {\bibfield  {journal} {\bibinfo  {journal} {Nature Materials}\ }\textbf {\bibinfo {volume} {19}},\ \bibinfo {pages} {170} (\bibinfo {year} {2020})}\BibitemShut {NoStop}%
\bibitem [{\citenamefont {Ingla-Aynés}\ \emph {et~al.}(2021)\citenamefont {Ingla-Aynés}, \citenamefont {Herling}, \citenamefont {Fabian}, \citenamefont {Hueso},\ and\ \citenamefont {Casanova}}]{ingla-aynes_electrical_2021}%
  \BibitemOpen
  \bibfield  {author} {\bibinfo {author} {\bibfnamefont {J.}~\bibnamefont {Ingla-Aynés}}, \bibinfo {author} {\bibfnamefont {F.}~\bibnamefont {Herling}}, \bibinfo {author} {\bibfnamefont {J.}~\bibnamefont {Fabian}}, \bibinfo {author} {\bibfnamefont {L.~E.}\ \bibnamefont {Hueso}}, \ and\ \bibinfo {author} {\bibfnamefont {F.}~\bibnamefont {Casanova}},\ }\href {\doibase 10.1103/PhysRevLett.127.047202} {\bibfield  {journal} {\bibinfo  {journal} {Physical Review Letters}\ }\textbf {\bibinfo {volume} {127}},\ \bibinfo {pages} {047202} (\bibinfo {year} {2021})}\BibitemShut {NoStop}%
\bibitem [{\citenamefont {Ingla-Aynés}\ \emph {et~al.}(2022)\citenamefont {Ingla-Aynés}, \citenamefont {Groen}, \citenamefont {Herling}, \citenamefont {Ontoso}, \citenamefont {Safeer}, \citenamefont {De~Juan}, \citenamefont {Hueso}, \citenamefont {Gobbi},\ and\ \citenamefont {Casanova}}]{ingla-aynes_omnidirectional_2022}%
  \BibitemOpen
  \bibfield  {author} {\bibinfo {author} {\bibfnamefont {J.}~\bibnamefont {Ingla-Aynés}}, \bibinfo {author} {\bibfnamefont {I.}~\bibnamefont {Groen}}, \bibinfo {author} {\bibfnamefont {F.}~\bibnamefont {Herling}}, \bibinfo {author} {\bibfnamefont {N.}~\bibnamefont {Ontoso}}, \bibinfo {author} {\bibfnamefont {C.~K.}\ \bibnamefont {Safeer}}, \bibinfo {author} {\bibfnamefont {F.}~\bibnamefont {De~Juan}}, \bibinfo {author} {\bibfnamefont {L.~E.}\ \bibnamefont {Hueso}}, \bibinfo {author} {\bibfnamefont {M.}~\bibnamefont {Gobbi}}, \ and\ \bibinfo {author} {\bibfnamefont {F.}~\bibnamefont {Casanova}},\ }\href {\doibase 10.1088/2053-1583/ac76d1} {\bibfield  {journal} {\bibinfo  {journal} {2D Materials}\ }\textbf {\bibinfo {volume} {9}},\ \bibinfo {pages} {045001} (\bibinfo {year} {2022})}\BibitemShut {NoStop}%
\bibitem [{\citenamefont {Safeer}\ \emph {et~al.}(2022)\citenamefont {Safeer}, \citenamefont {Herling}, \citenamefont {Choi}, \citenamefont {Ontoso}, \citenamefont {Ingla-Aynés}, \citenamefont {Hueso},\ and\ \citenamefont {Casanova}}]{safeer_reliability_2022}%
  \BibitemOpen
  \bibfield  {author} {\bibinfo {author} {\bibfnamefont {C.~K.}\ \bibnamefont {Safeer}}, \bibinfo {author} {\bibfnamefont {F.}~\bibnamefont {Herling}}, \bibinfo {author} {\bibfnamefont {W.~Y.}\ \bibnamefont {Choi}}, \bibinfo {author} {\bibfnamefont {N.}~\bibnamefont {Ontoso}}, \bibinfo {author} {\bibfnamefont {J.}~\bibnamefont {Ingla-Aynés}}, \bibinfo {author} {\bibfnamefont {L.~E.}\ \bibnamefont {Hueso}}, \ and\ \bibinfo {author} {\bibfnamefont {F.}~\bibnamefont {Casanova}},\ }\href {\doibase 10.1088/2053-1583/ac3c9b} {\bibfield  {journal} {\bibinfo  {journal} {2D Materials}\ }\textbf {\bibinfo {volume} {9}},\ \bibinfo {pages} {015024} (\bibinfo {year} {2022})}\BibitemShut {NoStop}%
\bibitem [{\citenamefont {Yang}\ \emph {et~al.}(2024{\natexlab{a}})\citenamefont {Yang}, \citenamefont {Chi}, \citenamefont {Avedissian}, \citenamefont {Dolan}, \citenamefont {Karuppasamy}, \citenamefont {Mart{\'\i}n-Garc{\'\i}a}, \citenamefont {Gobbi}, \citenamefont {Sofer}, \citenamefont {Hueso},\ and\ \citenamefont {Casanova}}]{yang_gatetunable_2024}%
  \BibitemOpen
  \bibfield  {author} {\bibinfo {author} {\bibfnamefont {H.}~\bibnamefont {Yang}}, \bibinfo {author} {\bibfnamefont {Z.}~\bibnamefont {Chi}}, \bibinfo {author} {\bibfnamefont {G.}~\bibnamefont {Avedissian}}, \bibinfo {author} {\bibfnamefont {E.}~\bibnamefont {Dolan}}, \bibinfo {author} {\bibfnamefont {M.}~\bibnamefont {Karuppasamy}}, \bibinfo {author} {\bibfnamefont {B.}~\bibnamefont {Mart{\'\i}n-Garc{\'\i}a}}, \bibinfo {author} {\bibfnamefont {M.}~\bibnamefont {Gobbi}}, \bibinfo {author} {\bibfnamefont {Z.}~\bibnamefont {Sofer}}, \bibinfo {author} {\bibfnamefont {L.~E.}\ \bibnamefont {Hueso}}, \ and\ \bibinfo {author} {\bibfnamefont {F.}~\bibnamefont {Casanova}},\ }\href@noop {} {\bibfield  {journal} {\bibinfo  {journal} {Adv. Funct. Mater.}\ }\textbf {\bibinfo {volume} {34}},\ \bibinfo {pages} {2404872} (\bibinfo {year} {2024}{\natexlab{a}})}\BibitemShut {NoStop}%
\bibitem [{\citenamefont {Yang}\ \emph {et~al.}(2024{\natexlab{b}})\citenamefont {Yang}, \citenamefont {Mart{\'\i}n-Garc{\'\i}a}, \citenamefont {Kim{\'a}k}, \citenamefont {Schmoranzerov{\'a}}, \citenamefont {Dolan}, \citenamefont {Chi}, \citenamefont {Gobbi}, \citenamefont {N{\v e}mec}, \citenamefont {Hueso},\ and\ \citenamefont {Casanova}}]{yang_twist-angle-tunable_2024}%
  \BibitemOpen
  \bibfield  {author} {\bibinfo {author} {\bibfnamefont {H.}~\bibnamefont {Yang}}, \bibinfo {author} {\bibfnamefont {B.}~\bibnamefont {Mart{\'\i}n-Garc{\'\i}a}}, \bibinfo {author} {\bibfnamefont {J.}~\bibnamefont {Kim{\'a}k}}, \bibinfo {author} {\bibfnamefont {E.}~\bibnamefont {Schmoranzerov{\'a}}}, \bibinfo {author} {\bibfnamefont {E.}~\bibnamefont {Dolan}}, \bibinfo {author} {\bibfnamefont {Z.}~\bibnamefont {Chi}}, \bibinfo {author} {\bibfnamefont {M.}~\bibnamefont {Gobbi}}, \bibinfo {author} {\bibfnamefont {P.}~\bibnamefont {N{\v e}mec}}, \bibinfo {author} {\bibfnamefont {L.~E.}\ \bibnamefont {Hueso}}, \ and\ \bibinfo {author} {\bibfnamefont {F.}~\bibnamefont {Casanova}},\ }\href@noop {} {\bibfield  {journal} {\bibinfo  {journal} {Nat. Mater.}\ }\textbf {\bibinfo {volume} {23}},\ \bibinfo {pages} {1502} (\bibinfo {year} {2024}{\natexlab{b}})}\BibitemShut {NoStop}%
\bibitem [{\citenamefont {Chi}\ \emph {et~al.}(2024)\citenamefont {Chi}, \citenamefont {Lee}, \citenamefont {Yang}, \citenamefont {Dolan}, \citenamefont {Safeer}, \citenamefont {Ingla‐Aynés}, \citenamefont {Herling}, \citenamefont {Ontoso}, \citenamefont {Martín‐García}, \citenamefont {Gobbi}, \citenamefont {Low}, \citenamefont {Hueso},\ and\ \citenamefont {Casanova}}]{chi_control_2024}%
  \BibitemOpen
  \bibfield  {author} {\bibinfo {author} {\bibfnamefont {Z.}~\bibnamefont {Chi}}, \bibinfo {author} {\bibfnamefont {S.}~\bibnamefont {Lee}}, \bibinfo {author} {\bibfnamefont {H.}~\bibnamefont {Yang}}, \bibinfo {author} {\bibfnamefont {E.}~\bibnamefont {Dolan}}, \bibinfo {author} {\bibfnamefont {C.~K.}\ \bibnamefont {Safeer}}, \bibinfo {author} {\bibfnamefont {J.}~\bibnamefont {Ingla‐Aynés}}, \bibinfo {author} {\bibfnamefont {F.}~\bibnamefont {Herling}}, \bibinfo {author} {\bibfnamefont {N.}~\bibnamefont {Ontoso}}, \bibinfo {author} {\bibfnamefont {B.}~\bibnamefont {Martín‐García}}, \bibinfo {author} {\bibfnamefont {M.}~\bibnamefont {Gobbi}}, \bibinfo {author} {\bibfnamefont {T.}~\bibnamefont {Low}}, \bibinfo {author} {\bibfnamefont {L.~E.}\ \bibnamefont {Hueso}}, \ and\ \bibinfo {author} {\bibfnamefont {F.}~\bibnamefont {Casanova}},\ }\href {\doibase 10.1002/adma.202310768} {\bibfield  {journal} {\bibinfo  {journal} {Advanced Materials}\ }\textbf {\bibinfo {volume} {36}},\ \bibinfo {pages} {2310768}
  (\bibinfo {year} {2024})}\BibitemShut {NoStop}%
\bibitem [{\citenamefont {Sinova}\ \emph {et~al.}(2015)\citenamefont {Sinova}, \citenamefont {Valenzuela}, \citenamefont {Wunderlich}, \citenamefont {Back},\ and\ \citenamefont {Jungwirth}}]{sinova_spin_2015}%
  \BibitemOpen
  \bibfield  {author} {\bibinfo {author} {\bibfnamefont {J.}~\bibnamefont {Sinova}}, \bibinfo {author} {\bibfnamefont {S.~O.}\ \bibnamefont {Valenzuela}}, \bibinfo {author} {\bibfnamefont {J.}~\bibnamefont {Wunderlich}}, \bibinfo {author} {\bibfnamefont {C.}~\bibnamefont {Back}}, \ and\ \bibinfo {author} {\bibfnamefont {T.}~\bibnamefont {Jungwirth}},\ }\href {\doibase 10.1103/RevModPhys.87.1213} {\bibfield  {journal} {\bibinfo  {journal} {Reviews of Modern Physics}\ }\textbf {\bibinfo {volume} {87}},\ \bibinfo {pages} {1213} (\bibinfo {year} {2015})}\BibitemShut {NoStop}%
\bibitem [{\citenamefont {Edelstein}(1990)}]{edelstein_spin_1990}%
  \BibitemOpen
  \bibfield  {author} {\bibinfo {author} {\bibfnamefont {V.}~\bibnamefont {Edelstein}},\ }\href {\doibase 10.1016/0038-1098(90)90963-C} {\bibfield  {journal} {\bibinfo  {journal} {Solid State Communications}\ }\textbf {\bibinfo {volume} {73}},\ \bibinfo {pages} {233} (\bibinfo {year} {1990})}\BibitemShut {NoStop}%
\bibitem [{\citenamefont {Wan}\ \emph {et~al.}(2022)\citenamefont {Wan}, \citenamefont {Hu}, \citenamefont {Mao}, \citenamefont {Fu}, \citenamefont {Yuan}, \citenamefont {Song}, \citenamefont {Gan}, \citenamefont {Xu}, \citenamefont {Xue}, \citenamefont {Cheng}, \citenamefont {Huang}, \citenamefont {Yang}, \citenamefont {Dai}, \citenamefont {Zeng},\ and\ \citenamefont {Kan}}]{wan_room-temperature_2022}%
  \BibitemOpen
  \bibfield  {author} {\bibinfo {author} {\bibfnamefont {Y.}~\bibnamefont {Wan}}, \bibinfo {author} {\bibfnamefont {T.}~\bibnamefont {Hu}}, \bibinfo {author} {\bibfnamefont {X.}~\bibnamefont {Mao}}, \bibinfo {author} {\bibfnamefont {J.}~\bibnamefont {Fu}}, \bibinfo {author} {\bibfnamefont {K.}~\bibnamefont {Yuan}}, \bibinfo {author} {\bibfnamefont {Y.}~\bibnamefont {Song}}, \bibinfo {author} {\bibfnamefont {X.}~\bibnamefont {Gan}}, \bibinfo {author} {\bibfnamefont {X.}~\bibnamefont {Xu}}, \bibinfo {author} {\bibfnamefont {M.}~\bibnamefont {Xue}}, \bibinfo {author} {\bibfnamefont {X.}~\bibnamefont {Cheng}}, \bibinfo {author} {\bibfnamefont {C.}~\bibnamefont {Huang}}, \bibinfo {author} {\bibfnamefont {J.}~\bibnamefont {Yang}}, \bibinfo {author} {\bibfnamefont {L.}~\bibnamefont {Dai}}, \bibinfo {author} {\bibfnamefont {H.}~\bibnamefont {Zeng}}, \ and\ \bibinfo {author} {\bibfnamefont {E.}~\bibnamefont {Kan}},\ }\href {\doibase 10.1103/PhysRevLett.128.067601} {\bibfield  {journal} {\bibinfo  {journal} {Physical
  Review Letters}\ }\textbf {\bibinfo {volume} {128}},\ \bibinfo {pages} {067601} (\bibinfo {year} {2022})}\BibitemShut {NoStop}%
\bibitem [{\citenamefont {Ontoso}\ \emph {et~al.}(2023{\natexlab{a}})\citenamefont {Ontoso}, \citenamefont {Safeer}, \citenamefont {Herling}, \citenamefont {Ingla-Aynés}, \citenamefont {Yang}, \citenamefont {Chi}, \citenamefont {Martin-Garcia}, \citenamefont {Robredo}, \citenamefont {Vergniory}, \citenamefont {De~Juan}, \citenamefont {Reyes~Calvo}, \citenamefont {Hueso},\ and\ \citenamefont {Casanova}}]{ontoso_unconventional_2023}%
  \BibitemOpen
  \bibfield  {author} {\bibinfo {author} {\bibfnamefont {N.}~\bibnamefont {Ontoso}}, \bibinfo {author} {\bibfnamefont {C.~K.}\ \bibnamefont {Safeer}}, \bibinfo {author} {\bibfnamefont {F.}~\bibnamefont {Herling}}, \bibinfo {author} {\bibfnamefont {J.}~\bibnamefont {Ingla-Aynés}}, \bibinfo {author} {\bibfnamefont {H.}~\bibnamefont {Yang}}, \bibinfo {author} {\bibfnamefont {Z.}~\bibnamefont {Chi}}, \bibinfo {author} {\bibfnamefont {B.}~\bibnamefont {Martin-Garcia}}, \bibinfo {author} {\bibfnamefont {I.}~\bibnamefont {Robredo}}, \bibinfo {author} {\bibfnamefont {M.~G.}\ \bibnamefont {Vergniory}}, \bibinfo {author} {\bibfnamefont {F.}~\bibnamefont {De~Juan}}, \bibinfo {author} {\bibfnamefont {M.}~\bibnamefont {Reyes~Calvo}}, \bibinfo {author} {\bibfnamefont {L.~E.}\ \bibnamefont {Hueso}}, \ and\ \bibinfo {author} {\bibfnamefont {F.}~\bibnamefont {Casanova}},\ }\href {\doibase 10.1103/PhysRevApplied.19.014053} {\bibfield  {journal} {\bibinfo  {journal} {Physical Review Applied}\ }\textbf {\bibinfo {volume}
  {19}},\ \bibinfo {pages} {014053} (\bibinfo {year} {2023}{\natexlab{a}})}\BibitemShut {NoStop}%
\bibitem [{\citenamefont {Camosi}\ \emph {et~al.}(2022)\citenamefont {Camosi}, \citenamefont {Sv{\v e}tl{\'\i}k}, \citenamefont {Costache}, \citenamefont {Savero~Torres}, \citenamefont {Fern{\'a}ndez~Aguirre}, \citenamefont {Marinova}, \citenamefont {Dimitrov}, \citenamefont {Gospodinov}, \citenamefont {Sierra},\ and\ \citenamefont {Valenzuela}}]{Camosi2022-yn}%
  \BibitemOpen
  \bibfield  {author} {\bibinfo {author} {\bibfnamefont {L.}~\bibnamefont {Camosi}}, \bibinfo {author} {\bibfnamefont {J.}~\bibnamefont {Sv{\v e}tl{\'\i}k}}, \bibinfo {author} {\bibfnamefont {M.~V.}\ \bibnamefont {Costache}}, \bibinfo {author} {\bibfnamefont {W.}~\bibnamefont {Savero~Torres}}, \bibinfo {author} {\bibfnamefont {I.}~\bibnamefont {Fern{\'a}ndez~Aguirre}}, \bibinfo {author} {\bibfnamefont {V.}~\bibnamefont {Marinova}}, \bibinfo {author} {\bibfnamefont {D.}~\bibnamefont {Dimitrov}}, \bibinfo {author} {\bibfnamefont {M.}~\bibnamefont {Gospodinov}}, \bibinfo {author} {\bibfnamefont {J.~F.}\ \bibnamefont {Sierra}}, \ and\ \bibinfo {author} {\bibfnamefont {S.~O.}\ \bibnamefont {Valenzuela}},\ }\href@noop {} {\bibfield  {journal} {\bibinfo  {journal} {2d Mater.}\ }\textbf {\bibinfo {volume} {9}},\ \bibinfo {pages} {035014} (\bibinfo {year} {2022})}\BibitemShut {NoStop}%
\bibitem [{\citenamefont {Savero~Torres}\ \emph {et~al.}(2017)\citenamefont {Savero~Torres}, \citenamefont {Sierra}, \citenamefont {Ben{\'\i}tez}, \citenamefont {Bonell}, \citenamefont {Costache},\ and\ \citenamefont {Valenzuela}}]{Savero_Torres2017-qk}%
  \BibitemOpen
  \bibfield  {author} {\bibinfo {author} {\bibfnamefont {W.}~\bibnamefont {Savero~Torres}}, \bibinfo {author} {\bibfnamefont {J.~F.}\ \bibnamefont {Sierra}}, \bibinfo {author} {\bibfnamefont {L.~A.}\ \bibnamefont {Ben{\'\i}tez}}, \bibinfo {author} {\bibfnamefont {F.}~\bibnamefont {Bonell}}, \bibinfo {author} {\bibfnamefont {M.~V.}\ \bibnamefont {Costache}}, \ and\ \bibinfo {author} {\bibfnamefont {S.~O.}\ \bibnamefont {Valenzuela}},\ }\href@noop {} {\bibfield  {journal} {\bibinfo  {journal} {2d Mater.}\ }\textbf {\bibinfo {volume} {4}},\ \bibinfo {pages} {041008} (\bibinfo {year} {2017})}\BibitemShut {NoStop}%
\bibitem [{\citenamefont {Ghiasi}\ \emph {et~al.}(2017)\citenamefont {Ghiasi}, \citenamefont {Ingla-Ayn{\'e}s}, \citenamefont {Kaverzin},\ and\ \citenamefont {van Wees}}]{Ghiasi2017-fa}%
  \BibitemOpen
  \bibfield  {author} {\bibinfo {author} {\bibfnamefont {T.~S.}\ \bibnamefont {Ghiasi}}, \bibinfo {author} {\bibfnamefont {J.}~\bibnamefont {Ingla-Ayn{\'e}s}}, \bibinfo {author} {\bibfnamefont {A.~A.}\ \bibnamefont {Kaverzin}}, \ and\ \bibinfo {author} {\bibfnamefont {B.~J.}\ \bibnamefont {van Wees}},\ }\href@noop {} {\bibfield  {journal} {\bibinfo  {journal} {Nano Lett.}\ }\textbf {\bibinfo {volume} {17}},\ \bibinfo {pages} {7528} (\bibinfo {year} {2017})}\BibitemShut {NoStop}%
\bibitem [{\citenamefont {Ben{\'\i}tez}\ \emph {et~al.}(2018)\citenamefont {Ben{\'\i}tez}, \citenamefont {Sierra}, \citenamefont {Savero~Torres}, \citenamefont {Arrighi}, \citenamefont {Bonell}, \citenamefont {Costache},\ and\ \citenamefont {Valenzuela}}]{Benitez2018-uc}%
  \BibitemOpen
  \bibfield  {author} {\bibinfo {author} {\bibfnamefont {L.~A.}\ \bibnamefont {Ben{\'\i}tez}}, \bibinfo {author} {\bibfnamefont {J.~F.}\ \bibnamefont {Sierra}}, \bibinfo {author} {\bibfnamefont {W.}~\bibnamefont {Savero~Torres}}, \bibinfo {author} {\bibfnamefont {A.}~\bibnamefont {Arrighi}}, \bibinfo {author} {\bibfnamefont {F.}~\bibnamefont {Bonell}}, \bibinfo {author} {\bibfnamefont {M.~V.}\ \bibnamefont {Costache}}, \ and\ \bibinfo {author} {\bibfnamefont {S.~O.}\ \bibnamefont {Valenzuela}},\ }\href@noop {} {\bibfield  {journal} {\bibinfo  {journal} {Nat. Phys.}\ }\textbf {\bibinfo {volume} {14}},\ \bibinfo {pages} {303} (\bibinfo {year} {2018})}\BibitemShut {NoStop}%
\bibitem [{\citenamefont {Sierra}\ \emph {et~al.}(2025)\citenamefont {Sierra}, \citenamefont {Sv{\v e}tl{\'\i}k}, \citenamefont {Savero~Torres}, \citenamefont {Camosi}, \citenamefont {Herling}, \citenamefont {Guillet}, \citenamefont {Xu}, \citenamefont {Reparaz}, \citenamefont {Marinova}, \citenamefont {Dimitrov},\ and\ \citenamefont {Valenzuela}}]{Sierra2025-tc}%
  \BibitemOpen
  \bibfield  {author} {\bibinfo {author} {\bibfnamefont {J.~F.}\ \bibnamefont {Sierra}}, \bibinfo {author} {\bibfnamefont {J.}~\bibnamefont {Sv{\v e}tl{\'\i}k}}, \bibinfo {author} {\bibfnamefont {W.}~\bibnamefont {Savero~Torres}}, \bibinfo {author} {\bibfnamefont {L.}~\bibnamefont {Camosi}}, \bibinfo {author} {\bibfnamefont {F.}~\bibnamefont {Herling}}, \bibinfo {author} {\bibfnamefont {T.}~\bibnamefont {Guillet}}, \bibinfo {author} {\bibfnamefont {K.}~\bibnamefont {Xu}}, \bibinfo {author} {\bibfnamefont {J.~S.}\ \bibnamefont {Reparaz}}, \bibinfo {author} {\bibfnamefont {V.}~\bibnamefont {Marinova}}, \bibinfo {author} {\bibfnamefont {D.}~\bibnamefont {Dimitrov}}, \ and\ \bibinfo {author} {\bibfnamefont {S.~O.}\ \bibnamefont {Valenzuela}},\ }\href@noop {} {\bibfield  {journal} {\bibinfo  {journal} {Nat. Mater.}\ }\textbf {\bibinfo {volume} {24}},\ \bibinfo {pages} {876} (\bibinfo {year} {2025})}\BibitemShut {NoStop}%
\bibitem [{\citenamefont {Ontoso}\ \emph {et~al.}(2023{\natexlab{b}})\citenamefont {Ontoso}, \citenamefont {Safeer}, \citenamefont {Ingla-Ayn{\'e}s}, \citenamefont {Herling}, \citenamefont {Hueso}, \citenamefont {Calvo},\ and\ \citenamefont {Casanova}}]{Ontoso2023-kr}%
  \BibitemOpen
  \bibfield  {author} {\bibinfo {author} {\bibfnamefont {N.}~\bibnamefont {Ontoso}}, \bibinfo {author} {\bibfnamefont {C.~K.}\ \bibnamefont {Safeer}}, \bibinfo {author} {\bibfnamefont {J.}~\bibnamefont {Ingla-Ayn{\'e}s}}, \bibinfo {author} {\bibfnamefont {F.}~\bibnamefont {Herling}}, \bibinfo {author} {\bibfnamefont {L.~E.}\ \bibnamefont {Hueso}}, \bibinfo {author} {\bibfnamefont {M.~R.}\ \bibnamefont {Calvo}}, \ and\ \bibinfo {author} {\bibfnamefont {F.}~\bibnamefont {Casanova}},\ }\href@noop {} {\bibfield  {journal} {\bibinfo  {journal} {Appl. Phys. Lett.}\ }\textbf {\bibinfo {volume} {123}},\ \bibinfo {pages} {032401} (\bibinfo {year} {2023}{\natexlab{b}})}\BibitemShut {NoStop}%
\bibitem [{\citenamefont {MacNeill}\ \emph {et~al.}(2017)\citenamefont {MacNeill}, \citenamefont {Stiehl}, \citenamefont {Guimaraes}, \citenamefont {Buhrman}, \citenamefont {Park},\ and\ \citenamefont {Ralph}}]{MacNeill2017-rd}%
  \BibitemOpen
  \bibfield  {author} {\bibinfo {author} {\bibfnamefont {D.}~\bibnamefont {MacNeill}}, \bibinfo {author} {\bibfnamefont {G.~M.}\ \bibnamefont {Stiehl}}, \bibinfo {author} {\bibfnamefont {M.~H.~D.}\ \bibnamefont {Guimaraes}}, \bibinfo {author} {\bibfnamefont {R.~A.}\ \bibnamefont {Buhrman}}, \bibinfo {author} {\bibfnamefont {J.}~\bibnamefont {Park}}, \ and\ \bibinfo {author} {\bibfnamefont {D.~C.}\ \bibnamefont {Ralph}},\ }\href@noop {} {\bibfield  {journal} {\bibinfo  {journal} {Nat. Phys.}\ }\textbf {\bibinfo {volume} {13}},\ \bibinfo {pages} {300} (\bibinfo {year} {2017})}\BibitemShut {NoStop}%
\bibitem [{\citenamefont {Lee}\ \emph {et~al.}(2022)\citenamefont {Lee}, \citenamefont {De~Sousa}, \citenamefont {Kwon}, \citenamefont {De~Juan}, \citenamefont {Chi}, \citenamefont {Casanova},\ and\ \citenamefont {Low}}]{lee_charge--spin_2022}%
  \BibitemOpen
  \bibfield  {author} {\bibinfo {author} {\bibfnamefont {S.}~\bibnamefont {Lee}}, \bibinfo {author} {\bibfnamefont {D.~J.~P.}\ \bibnamefont {De~Sousa}}, \bibinfo {author} {\bibfnamefont {Y.-K.}\ \bibnamefont {Kwon}}, \bibinfo {author} {\bibfnamefont {F.}~\bibnamefont {De~Juan}}, \bibinfo {author} {\bibfnamefont {Z.}~\bibnamefont {Chi}}, \bibinfo {author} {\bibfnamefont {F.}~\bibnamefont {Casanova}}, \ and\ \bibinfo {author} {\bibfnamefont {T.}~\bibnamefont {Low}},\ }\href {\doibase 10.1103/PhysRevB.106.165420} {\bibfield  {journal} {\bibinfo  {journal} {Physical Review B}\ }\textbf {\bibinfo {volume} {106}},\ \bibinfo {pages} {165420} (\bibinfo {year} {2022})}\BibitemShut {NoStop}%
\bibitem [{\citenamefont {Yan}\ \emph {et~al.}(2016)\citenamefont {Yan}, \citenamefont {Txoperena}, \citenamefont {Llopis}, \citenamefont {Dery}, \citenamefont {Hueso},\ and\ \citenamefont {Casanova}}]{yan_two-dimensional_2016}%
  \BibitemOpen
  \bibfield  {author} {\bibinfo {author} {\bibfnamefont {W.}~\bibnamefont {Yan}}, \bibinfo {author} {\bibfnamefont {O.}~\bibnamefont {Txoperena}}, \bibinfo {author} {\bibfnamefont {R.}~\bibnamefont {Llopis}}, \bibinfo {author} {\bibfnamefont {H.}~\bibnamefont {Dery}}, \bibinfo {author} {\bibfnamefont {L.~E.}\ \bibnamefont {Hueso}}, \ and\ \bibinfo {author} {\bibfnamefont {F.}~\bibnamefont {Casanova}},\ }\href {\doibase 10.1038/ncomms13372} {\bibfield  {journal} {\bibinfo  {journal} {Nature Communications}\ }\textbf {\bibinfo {volume} {7}},\ \bibinfo {pages} {13372} (\bibinfo {year} {2016})}\BibitemShut {NoStop}%
\bibitem [{\citenamefont {Garcia}\ \emph {et~al.}(2018)\citenamefont {Garcia}, \citenamefont {Vila}, \citenamefont {Cummings},\ and\ \citenamefont {Roche}}]{garcia_spin_2018}%
  \BibitemOpen
  \bibfield  {author} {\bibinfo {author} {\bibfnamefont {J.~H.}\ \bibnamefont {Garcia}}, \bibinfo {author} {\bibfnamefont {M.}~\bibnamefont {Vila}}, \bibinfo {author} {\bibfnamefont {A.~W.}\ \bibnamefont {Cummings}}, \ and\ \bibinfo {author} {\bibfnamefont {S.}~\bibnamefont {Roche}},\ }\href {\doibase 10.1039/C7CS00864C} {\bibfield  {journal} {\bibinfo  {journal} {Chemical Society Reviews}\ }\textbf {\bibinfo {volume} {47}},\ \bibinfo {pages} {3359} (\bibinfo {year} {2018})}\BibitemShut {NoStop}%
\bibitem [{\citenamefont {Garcia}, \citenamefont {Cummings},\ and\ \citenamefont {Roche}(2017)}]{garcia_spin_2017}%
  \BibitemOpen
  \bibfield  {author} {\bibinfo {author} {\bibfnamefont {J.~H.}\ \bibnamefont {Garcia}}, \bibinfo {author} {\bibfnamefont {A.~W.}\ \bibnamefont {Cummings}}, \ and\ \bibinfo {author} {\bibfnamefont {S.}~\bibnamefont {Roche}},\ }\href {\doibase 10.1021/acs.nanolett.7b02364} {\bibfield  {journal} {\bibinfo  {journal} {Nano Letters}\ }\textbf {\bibinfo {volume} {17}},\ \bibinfo {pages} {5078} (\bibinfo {year} {2017})}\BibitemShut {NoStop}%
\end{thebibliography}

\begin{thebibliography}{5}%
\makeatletter
\providecommand \@ifxundefined [1]{%
 \@ifx{#1\undefined}
}%
\providecommand \@ifnum [1]{%
 \ifnum #1\expandafter \@firstoftwo
 \else \expandafter \@secondoftwo
 \fi
}%
\providecommand \@ifx [1]{%
 \ifx #1\expandafter \@firstoftwo
 \else \expandafter \@secondoftwo
 \fi
}%
\providecommand \natexlab [1]{#1}%
\providecommand \enquote  [1]{``#1''}%
\providecommand \bibnamefont  [1]{#1}%
\providecommand \bibfnamefont [1]{#1}%
\providecommand \citenamefont [1]{#1}%
\providecommand \href@noop [0]{\@secondoftwo}%
\providecommand \href [0]{\begingroup \@sanitize@url \@href}%
\providecommand \@href[1]{\@@startlink{#1}\@@href}%
\providecommand \@@href[1]{\endgroup#1\@@endlink}%
\providecommand \@sanitize@url [0]{\catcode `\\12\catcode `\$12\catcode `\&12\catcode `\#12\catcode `\^12\catcode `\_12\catcode `\%12\relax}%
\providecommand \@@startlink[1]{}%
\providecommand \@@endlink[0]{}%
\providecommand \url  [0]{\begingroup\@sanitize@url \@url }%
\providecommand \@url [1]{\endgroup\@href {#1}{\urlprefix }}%
\providecommand \urlprefix  [0]{URL }%
\providecommand \Eprint [0]{\href }%
\providecommand \doibase [0]{http://dx.doi.org/}%
\providecommand \selectlanguage [0]{\@gobble}%
\providecommand \bibinfo  [0]{\@secondoftwo}%
\providecommand \bibfield  [0]{\@secondoftwo}%
\providecommand \translation [1]{[#1]}%
\providecommand \BibitemOpen [0]{}%
\providecommand \bibitemStop [0]{}%
\providecommand \bibitemNoStop [0]{.\EOS\space}%
\providecommand \EOS [0]{\spacefactor3000\relax}%
\providecommand \BibitemShut  [1]{\csname bibitem#1\endcsname}%
\let\auto@bib@innerbib\@empty
\bibitem [{\citenamefont {Zhou}\ \emph {et~al.}(2024)\citenamefont {Zhou}, \citenamefont {Lu}, \citenamefont {Yang}, \citenamefont {Zhang}, \citenamefont {Liu}, \citenamefont {Zeng}, \citenamefont {Yan}, \citenamefont {Li}, \citenamefont {Wei}, \citenamefont {Wu}, \citenamefont {Pu}, \citenamefont {Liu}, \citenamefont {He}, \citenamefont {Zhang},\ and\ \citenamefont {Xu}}]{Zhou2024-tjAAA}%
  \BibitemOpen
  \bibfield  {author} {\bibinfo {author} {\bibfnamefont {J.}~\bibnamefont {Zhou}}, \bibinfo {author} {\bibfnamefont {X.}~\bibnamefont {Lu}}, \bibinfo {author} {\bibfnamefont {J.}~\bibnamefont {Yang}}, \bibinfo {author} {\bibfnamefont {X.}~\bibnamefont {Zhang}}, \bibinfo {author} {\bibfnamefont {Q.}~\bibnamefont {Liu}}, \bibinfo {author} {\bibfnamefont {Q.}~\bibnamefont {Zeng}}, \bibinfo {author} {\bibfnamefont {Y.}~\bibnamefont {Yan}}, \bibinfo {author} {\bibfnamefont {Y.}~\bibnamefont {Li}}, \bibinfo {author} {\bibfnamefont {L.}~\bibnamefont {Wei}}, \bibinfo {author} {\bibfnamefont {J.}~\bibnamefont {Wu}}, \bibinfo {author} {\bibfnamefont {Y.}~\bibnamefont {Pu}}, \bibinfo {author} {\bibfnamefont {R.}~\bibnamefont {Liu}}, \bibinfo {author} {\bibfnamefont {L.}~\bibnamefont {He}}, \bibinfo {author} {\bibfnamefont {R.}~\bibnamefont {Zhang}}, \ and\ \bibinfo {author} {\bibfnamefont {Y.}~\bibnamefont {Xu}},\ }\href@noop {} {\bibfield  {journal} {\bibinfo  {journal} {Carbon N. Y.}\ }\textbf {\bibinfo {volume}
  {228}},\ \bibinfo {pages} {119321} (\bibinfo {year} {2024})}\BibitemShut {NoStop}%
\bibitem [{\citenamefont {Benítez}\ \emph {et~al.}(2020)\citenamefont {Benítez}, \citenamefont {Savero~Torres}, \citenamefont {Sierra}, \citenamefont {Timmermans}, \citenamefont {Garcia}, \citenamefont {Roche}, \citenamefont {Costache},\ and\ \citenamefont {Valenzuela}}]{benitez_tunable_2020AAA}%
  \BibitemOpen
  \bibfield  {author} {\bibinfo {author} {\bibfnamefont {L.~A.}\ \bibnamefont {Benítez}}, \bibinfo {author} {\bibfnamefont {W.}~\bibnamefont {Savero~Torres}}, \bibinfo {author} {\bibfnamefont {J.~F.}\ \bibnamefont {Sierra}}, \bibinfo {author} {\bibfnamefont {M.}~\bibnamefont {Timmermans}}, \bibinfo {author} {\bibfnamefont {J.~H.}\ \bibnamefont {Garcia}}, \bibinfo {author} {\bibfnamefont {S.}~\bibnamefont {Roche}}, \bibinfo {author} {\bibfnamefont {M.~V.}\ \bibnamefont {Costache}}, \ and\ \bibinfo {author} {\bibfnamefont {S.~O.}\ \bibnamefont {Valenzuela}},\ }\href {\doibase 10.1038/s41563-019-0575-1} {\bibfield  {journal} {\bibinfo  {journal} {Nature Materials}\ }\textbf {\bibinfo {volume} {19}},\ \bibinfo {pages} {170} (\bibinfo {year} {2020})}\BibitemShut {NoStop}%
\bibitem [{\citenamefont {Safeer}\ \emph {et~al.}(2020)\citenamefont {Safeer}, \citenamefont {Ingla-Aynés}, \citenamefont {Ontoso}, \citenamefont {Herling}, \citenamefont {Yan}, \citenamefont {Hueso},\ and\ \citenamefont {Casanova}}]{safeer_spin_2020AAA}%
  \BibitemOpen
  \bibfield  {author} {\bibinfo {author} {\bibfnamefont {C.~K.}\ \bibnamefont {Safeer}}, \bibinfo {author} {\bibfnamefont {J.}~\bibnamefont {Ingla-Aynés}}, \bibinfo {author} {\bibfnamefont {N.}~\bibnamefont {Ontoso}}, \bibinfo {author} {\bibfnamefont {F.}~\bibnamefont {Herling}}, \bibinfo {author} {\bibfnamefont {W.}~\bibnamefont {Yan}}, \bibinfo {author} {\bibfnamefont {L.~E.}\ \bibnamefont {Hueso}}, \ and\ \bibinfo {author} {\bibfnamefont {F.}~\bibnamefont {Casanova}},\ }\href {\doibase 10.1021/acs.nanolett.0c01428} {\bibfield  {journal} {\bibinfo  {journal} {Nano Letters}\ }\textbf {\bibinfo {volume} {20}},\ \bibinfo {pages} {4573} (\bibinfo {year} {2020})}\BibitemShut {NoStop}%
\bibitem [{\citenamefont {Herling}\ \emph {et~al.}(2020)\citenamefont {Herling}, \citenamefont {Safeer}, \citenamefont {Ingla-Aynés}, \citenamefont {Ontoso}, \citenamefont {Hueso},\ and\ \citenamefont {Casanova}}]{herling_gate_2020AAA}%
  \BibitemOpen
  \bibfield  {author} {\bibinfo {author} {\bibfnamefont {F.}~\bibnamefont {Herling}}, \bibinfo {author} {\bibfnamefont {C.~K.}\ \bibnamefont {Safeer}}, \bibinfo {author} {\bibfnamefont {J.}~\bibnamefont {Ingla-Aynés}}, \bibinfo {author} {\bibfnamefont {N.}~\bibnamefont {Ontoso}}, \bibinfo {author} {\bibfnamefont {L.~E.}\ \bibnamefont {Hueso}}, \ and\ \bibinfo {author} {\bibfnamefont {F.}~\bibnamefont {Casanova}},\ }\href {\doibase 10.1063/5.0006101} {\bibfield  {journal} {\bibinfo  {journal} {APL Materials}\ }\textbf {\bibinfo {volume} {8}},\ \bibinfo {pages} {071103} (\bibinfo {year} {2020})}\BibitemShut {NoStop}%
\bibitem [{\citenamefont {Rashba}(2000)}]{rashba_theory_2000AAA}%
  \BibitemOpen
  \bibfield  {author} {\bibinfo {author} {\bibfnamefont {E.~I.}\ \bibnamefont {Rashba}},\ }\href {\doibase 10.1103/PhysRevB.62.R16267} {\bibfield  {journal} {\bibinfo  {journal} {Physical Review B}\ }\textbf {\bibinfo {volume} {62}},\ \bibinfo {pages} {R16267} (\bibinfo {year} {2000})}\BibitemShut {NoStop}%
\end{thebibliography}
\end{document}